\documentclass[preprint,12pt]{elsarticle}

\usepackage{tabularx}
\usepackage{xcolor,colortbl}
\usepackage{lipsum}
\usepackage{booktabs}
\usepackage{xstring}
\usepackage{comment}
\usepackage{varwidth}
\usepackage{enumitem}
\usepackage{wrapfig}
\usepackage{comment}
\usepackage{multirow}
\usepackage{caption}
\usepackage{makecell}
\usepackage{url}
\usepackage{subcaption}
\usepackage[normalem]{ulem}

\definecolor{codegreen}{rgb}{0,0.6,0}
\definecolor{codegray}{rgb}{0.5,0.5,0.5}
\definecolor{codepurple}{rgb}{0.58,0,0.82}
\definecolor{backcolour}{rgb}{0.95,0.95,0.92}
\definecolor{codekeyword}{RGB}{153, 51, 153}

\usepackage[frozencache,cachedir=.]{minted}
\setminted[text]{
    bgcolor=backcolour,
    fontsize=\scriptsize,
    baselinestretch=1,
    linenos
}

\usepackage{listings}

\usepackage{changepage}
\newcommand{\interviewquote}[2]{
 \def\FrameCommand{
    \hspace{0pt} 
    {\color{cyan}  \vrule width 2pt}%
    {\color{white} \vrule width 3pt}%
  }%
  \MakeFramed{\advance\hsize-\width\FrameRestore}%
  \noindent
  \begin{adjustwidth}{}{1pt}
  {\footnotesize``#1'' - {#2}}
  \end{adjustwidth}
  \endMakeFramed%
}

\usepackage[framemethod=TikZ]{mdframed}

\usepackage{listings}
\definecolor{codegreen}{rgb}{0,0.6,0}
\definecolor{codegray}{rgb}{0.5,0.5,0.5}
\definecolor{codepurple}{rgb}{0.969, 0.604, 0.827}
\definecolor{backcolour}{rgb}{0.95,0.95,0.92}

\lstdefinestyle{lststyle}{
    backgroundcolor=\color{backcolour},   
    commentstyle=\color{codegreen},
    keywordstyle=\color{magenta},
    numberstyle=\tiny\color{codegray},
    stringstyle=\color{codepurple},
    basicstyle=\ttfamily\footnotesize,
    breakatwhitespace=false,         
    breaklines=true,                 
    captionpos=b,                    
    keepspaces=true,                 
    numbers=left,                    
    numbersep=5pt,                  
    showspaces=false,                
    showstringspaces=false,
    showtabs=false,                  
    tabsize=2
}

\newenvironment{coloredframe}[2][]{
    \mdfsetup{
        linewidth=0pt,
        skipabove=3pt, skipbelow=3pt,% pads top.
        innerlinewidth=1.5pt, 
        innerlinecolor=white, % size of bar
        linewidth=0pt,
        backgroundcolor=#2!10,
        innerleftmargin=5pt, % Padding on the left
        innerrightmargin=10pt,      % Padding on the right
        innertopmargin=5pt,        % Padding at the top
        innerbottommargin=5pt,     % 
        roundcorner=5pt           % Radius of the round 
        }
    \begin{mdframed}}
    {\end{mdframed}}

\journal{Journal of Systems and Software}

\begin{document}

\begin{frontmatter}

\title{Exploring the Integration of Large Language Models in Industrial Test Maintenance Processes}

\author[label1,label2]{Jingxiong Liu}
\author[label1,label2]{Ludvig Lemner}
\author[label1,label2]{Linnea Wahlgren}
\author[label1]{Gregory Gay\corref{mycorrespondingauthor}}
\cortext[mycorrespondingauthor]{Corresponding author}
\ead{greg@greggay.com}
\author[label2,label1]{Nasser Mohammadiha}
\author[label2]{Joakim Wennerberg}
            
\affiliation[label1]{organization={Chalmers University of Technology and University of Gothenburg},
            % addressline={},
            city={Gothenburg},
            % postcode={},
            % state={},
            country={Sweden}}

\affiliation[label2]{organization={Ericsson AB},
            % addressline={},
            city={Gothenburg},
            % postcode={},
            % state={},
            country={Sweden}}   

\begin{abstract}

Much of the cost and effort required during the software testing process is invested in performing test maintenance---the addition, removal, or modification of test cases to keep the test suite in sync with the system-under-test or to otherwise improve its quality. Tool support could reduce the cost---and improve the quality---of test maintenance by automating aspects of the process or by providing guidance and support to developers. 

In this study, we explore the capabilities and applications of large language models (LLMs)---complex machine learning models adapted to textual analysis---to support test maintenance. We conducted a case study at Ericsson AB where we explore the \textit{triggers} that indicate the need for test maintenance, the \textit{actions} that LLMs can take, and the \textit{considerations} that must be made when deploying LLMs in an industrial setting. We also propose and demonstrate a multi-agent architecture that can predict which tests require maintenance following a change to the source code. Collectively, these contributions advance our theoretical and practical understanding of how LLMs can be deployed to benefit industrial test maintenance processes.

\end{abstract}

%%Graphical abstract
\begin{graphicalabstract}
\centering
    \includegraphics[width=\textwidth]{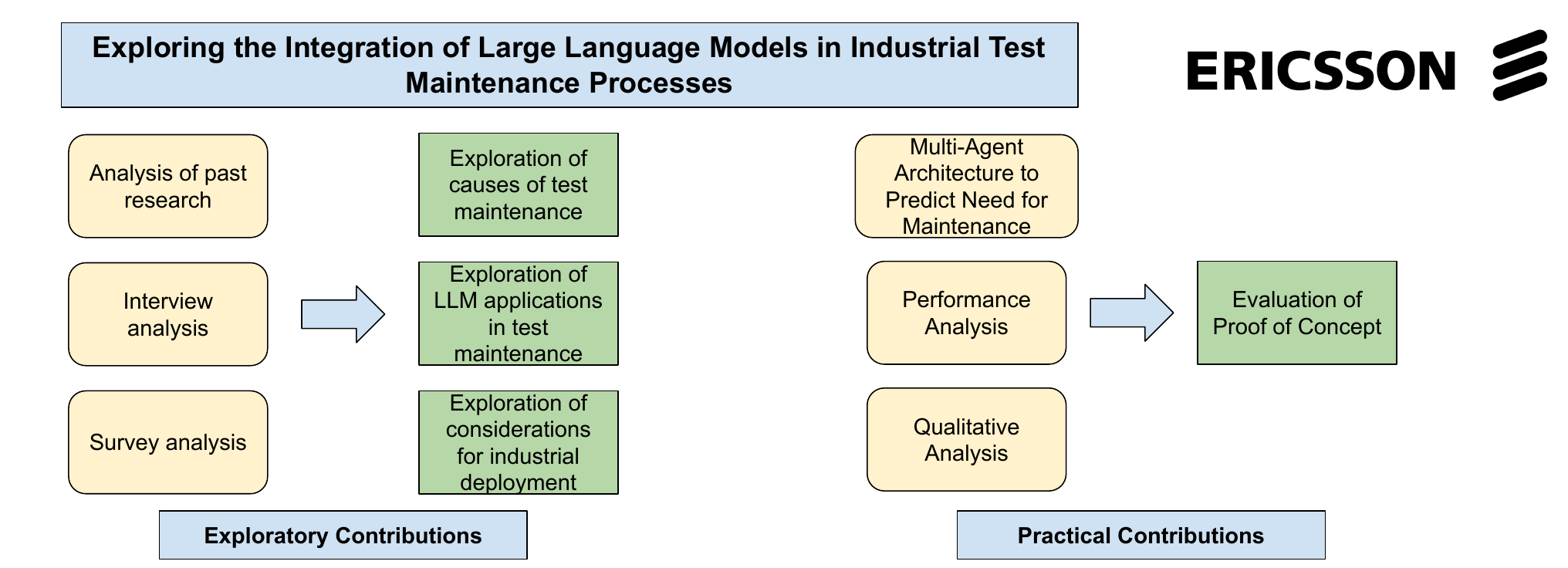}
\end{graphicalabstract}

% %%Research highlights, each must be less than 85 characters in total
\begin{highlights}
\item Research highlight 1: Industrial case study on use of LLMs in test maintenance. 
\item Research highlight 2: Explored maintenance triggers, actions LLMs can take, and deployment considerations.
\item Research highlight 3: Multi-agent architecture that predicts which tests need maintenance.
\item Research highlight 4: Two prototypes evaluated quantitatively and qualitatively on industrial codebase. 

\end{highlights}

\begin{keyword}
Software Testing, Test Maintenance, Machine Learning, Large Language Models, LLM Agent
\end{keyword}

\end{frontmatter}

\section{Introduction}\label{sec:intro}
Software testing---the application of selected input to a system-under-test and inspection of the resulting behavior---is a crucial component of the development process. However, it is also a notoriously expensive component, said to represent up to half of the total development cost of the system~\cite{alegroth2016maintenance}. 

Although there is an initial cost associated with creating test cases, much more cost is imposed by the need for \textit{test maintenance} as the system-under-test evolves~\cite{sneed2004cost}. During test maintenance, new tests are created and obsolete tests are modified or removed to adapt to changes in the evolving source code~\cite{alegroth2016maintenance}. Test maintenance can also be conducted to improve the quality or efficiency of the test suite---for example, to improve test coverage~\cite{sneed2004cost}. As a result, test maintenance requires long-term commitment of significant effort from software developers~\cite{sneed2004cost,wang2021understanding}.

Software tools have a role to play in reducing the cost and improving the quality or reliability of the test maintenance process~\cite{wang2021understanding,umar2019automation}. In particular, we focus on two forms of tool support. First, we examine cases where tools could automate certain tasks conducted during the maintenance process, such as the creation or modification of test code. Second, we consider scenarios where tools can offer support, such as examples or targeted suggestions, to human developers. Despite its importance in practice~\cite{kochhar2019practitioners}, test maintenance is an under-explored and poorly understood aspect of the overall testing process~\cite{gonzalez2017large,imtiaz2019systematic}. In relation to topics such as test generation, there has been comparatively little research on test maintenance automation~\cite{hu2023identify}. 

Large language models (LLMs), machine learning models trained on massive corpora of text, are an emerging technology with great potential for language analysis and transformation tasks such as translation, summarizing, and decision support due to their ability to infer semantic meaning from text~\cite{zhao2023survey}. These capabilities extend to both natural language and software code~\cite{zheng2023survey}. As a result, LLMs have shown promise in automating or assisting with software engineering~\cite{zhang2023survey} and software testing~\cite{wang2024software} tasks, including test  generation~\cite{schafer2023empirical}, documentation~\cite{hadi2023large}, refactoring~\cite{white2023chatgpt}, and automated program repair~\cite{sobania2023analysis}. LLMs have shown potential in both forms of tools support mentioned above, including automation~\cite{white2023chatgpt, surameery2023use} and serving as a conversational assistant to developers ~\cite{ross2023programmer}.

Although there has been significant interest in the general use of LLMs in software testing~\cite{wang2024software}, the applications of LLMs in the test maintenance process have only begun to be explored~\cite{chi2024reaccept,hu2023identify,liu2024fix}. Therefore, our purpose in this study is to examine the integration of LLMs and LLM agents---autonomous systems coupling an LLM with tool access and memory mechanisms~\cite{feldt2023towards}---into this process. We explore this topic through an industrial case study at Ericsson AB, a Swedish telecommunications company, with the aim of offering both exploratory and practical contributions to the field. 

The exploratory contributions of this work consist of an investigation of (a) the \textit{triggers} that indicate the potential need for test maintenance, (b) the potential \textit{actions} that LLMs or LLM agents can take based on those triggers, and (c), the \textit{considerations} that must be made when deploying LLMs in industry. To demonstrate the practical applicability of LLM agents in test maintenance, we also propose a multi-agent architecture that predicts which test cases require maintenance following a change to the source code. We evaluate two implementations of this architecture on an industrial codebase. Our quantitative and qualitative assessment of the prototypes offers advice for the future implementation of LLM agents in test maintenance. 

These contributions benefit both researchers and practitioners, advancing our understanding of how LLMs and LLM agents could be applied within test maintenance and offering practical guidance to industrial developers. 

The structure of this paper is as follows. In Section~\ref{sec:background}, we explain core background concepts. In Section~\ref{sec:related}, we explore related work. In Section~\ref{sec:methods}, we explain our research questions and the methods used to address these questions. In Section~\ref{sec:results}, we provide the results of our research activities. In Section~\ref{sec:discussion}, we answer the research questions and discuss threats to validity. Finally, in Section~\ref{sec:conclusions}, we offer conclusions. 

\section{Background}\label{sec:background}
\subsection{Software Testing}\label{sec:stbg}

\textit{Software testing} is an activity performed to ensure that software delivers correct functionality and meets quality goals (e.g., performance)~\cite{ISO/IEC/IEEE24765:2017}. During  testing, a \textit{test suite}---one or more \textit{test cases}---is executed against the system-under-test~\cite{ISO/IEC/IEEE24765:2017}. Each test case performs input actions, then compares the resulting program behavior against expectations (\textit{test oracles}~\cite{barr2014oracle}). 

%Testing can take place at multiple levels of granularity, including \textit{unit testing}---where individual pieces of code (e.g., a class) are tested in isolation~\cite{myers2004art}---\textit{integration testing}---where a set of cooperating units are tested~\cite{leung1990study}---and \textit{system testing}---where the full system is tested through a defined interface~\cite{myers2004art}. 

Software testing is an important, but notoriously expensive, activity~\cite{myers2004art}, especially due to the significant manual effort involved in test case creation or manual test execution~\cite{wang2021understanding}. Automation has a role to play in reducing the cost by, e.g.,  enabling automated execution~\cite{umar2019automation}. In particular, significant research efforts have been invested in automating the generation of test input~\cite{anand2013orchestrated}, including recent investigations of the capabilities of LLMs~\cite{wang2024software,yuan2023no}. %However, there are still significant concerns about the readability and maintainability of generated tests~\cite{palomba2016diffusion, grano2018empirical}. In addition, any form of test automation requires significant up-front effort to implement and additional effort to maintain~\cite{alegroth2016maintenance,garousi2016developing}.

\textit{Test maintenance} is the process of updating the test suite as the source code evolves---including, e.g., adding new tests for new functionality, removing tests that are no longer relevant, and adapting tests to ensure their continued relevance~\cite{alegroth2016maintenance}. The test suite may also evolve to improve its efficiency, e.g., by removing redundant tests, or to improve its thoroughness, e.g., by covering more diverse input or more paths through the codebase~\cite{sneed2004cost}. Test maintenance is understood to be important, and can account for a significant proportion of testing effort over the lifespan of a project~\cite{alegroth2016maintenance,sneed2004cost}. However, the area has received comparatively little research attention~\cite{skoglund2004case}. For example, ``test smells'' negatively impact test maintainability, but are significantly less well-understood than code smells~\cite{tufano2016empirical}.

\subsection{Generative Artificial Intelligence and Large-Language Models}\label{sec:genaibg}

Generative AI refers to machine learning models that produce content, e.g. text or images, in response to instructions delivered as a \textit{prompt}~\cite{cao2023comprehensive}. Prompts can include text and other media. Models are trained to infer the semantic meaning of a prompt, and use that meaning to produce an appropriate response. 

Current Generative AI models  are implemented using a transformer architecture, which is based on mechanisms designed to imitate the cognitive attention (i.e., the ability to focus on select stimuli) seen in humans~\cite{vaswani2017attention}. \textit{Foundation models} are trained on large and diverse data sets with the intention that they perform at a minimal level on new tasks~\cite{han2021pre}. These models can then be \textit{fine-tuned} using additional domain-specific training data~\cite{bommasani2021opportunities}.

\textit{Large language models (LLMs)} are a form of Generative AI suited for language analysis and transformation tasks such as translation, summarization, and decision support~\cite{achiam2023gpt}. LLMs iteratively predict the next element to add to the generated output~\cite{zhao2023survey}. Because of their ability to understand and output both natural language and code, LLMs are well-suited for software development tasks including test generation, automated program repair, and code review~\cite{wang2024software,ross2023programmer,
hou2023large, zhang2023unifying}.

An LLM and a human may not draw the same meaning from a prompt, which has led to the development of different prompting strategies, including \textit{few-shot prompting}---where input-output examples are supplied along with the prompt~\cite{brown2020language}---and \textit{chain-of-thought prompting}---where examples are augmented with reasoning explaining how the correct output was reached~\cite{wei2022chain}. 

LLMs are known to \textit{hallucinate} at times, issuing output that appears initially plausible, but is incorrect~\cite{ji2023survey}. Additionally, there is the risk of \textit{data leakage}, where the LLM may have seen a benchmark during training, leading to artificially high performance  on that benchmark~\cite{wang2024software}. 
%Applying LLMs in industrial applications is also not without its share of problems.
%Organisations may shy away from using commercial LLMs due to data privacy concerns and would prefer to use open-source or open-weight models instead. In any case, a general-purpose foundation model may not perform well within the context of the company, which can require a time and effort-intensive fine-tuning process. Further, constraints on computational power and energy consumption can limit the size of the model that an organization can reasonably utilize~\cite{wang2024software}.

An \textit{LLM agent} is a system coupling an LLM with tool access and memory capabilities~\cite{feldt2023towards}. These mechanisms allow LLM agents to reason, plan, and perceive or interact with an environment~\cite{feldt2023towards}. For example, an LLM agent may have access to a version control system or issue tracker through \textit{Retrieval-Augmented Generation} (RAG)~\cite{lewis2020retrieval}. This would allow it to reason about the code, make changes to it, and push those changes. Multiple agents can work together, each given different roles or sub-tasks~\cite{wu2023autogen}.

%A common tool built into LLM agents is \textit{Retrieval-Augmented Generation} (RAG)~\cite{lewis2020retrieval}. RAG allows an LLM to access information from external sources, and use that information to better respond to prompts. For example, even if a particular source code file was not part of the training data for an LLM, RAG could be used to retrieve that file and make inferences from its contents. 

\section{Related Work}\label{sec:related}

Maintainability is one of the characteristics of good test cases most prized by testers~\cite{kochhar2019practitioners}. However, in practice, few open-source projects implement tests following patterns that promote maintainability~\cite{gonzalez2017large}. Imtiaz et al. also noted a need for studies on test maintenance in  industry~\cite{imtiaz2019systematic}.

Multiple authors have examined factors that influence the probability, 
frequency, and difficulty of test maintenance. For example, Pinto et al. found that many of the tests ``added'' to suites are older tests that have been updated and renamed, that tests are generally deleted because they are obsolete and not because they are difficult to update, and that when tests are added, it is to cover new functionality or to validate bug fixes and refactored code~\cite{pinto2012understanding}. Alégroth et al. identified thirteen factors that affect the difficulty of test maintenance for GUI-based test suites~\cite{alegroth2016maintenance}, with test length having the greatest impact. Berglund et al. also identified factors that affect test maintenance for machine learning systems---e.g., non-determinism and explainability of the system-under-test---as well as factors that affect both traditional and machine learning systems---e.g., the required precision of the test oracle and consistency between teams~\cite{berglund2023test}. 

Researchers have examined co-evolution of test and source code, finding that changes to both source and test code within a short timeframe~\cite{sun2023revisiting} or shared naming conventions~\cite{chi2024reaccept,hu2023identify} often indicate co-evolution. Metrics such as Tconf can also assess the extent source and test code have successfully co-evolved~\cite{kitai2022have}.  Visualizations can also show how source and test code have evolved~\cite{ens2014chronotwigger}. Traceability links can also be inferred between source and test code based on co-evolution~\cite{sohn2022cement}.

Many researchers have worked to identify changes to source code that trigger a need for maintenance. We draw on this literature as part of answering our first research question, and discuss the studies in Section~\ref{sec:rq1_review_results}. 
%For example, Shimmi and Rahimi extracted patterns of co-evolution, classified as triggering addition, deletion, or modification of test cases~\cite{shimmi2022patterns}. Reich and Maalej extracted similar patterns, dividing code changes into low-level  (changes to a single file, e.g., a change to a variable's data type) and high-level changes to the project (e.g., merging two packages)~\cite{reich2023testability}. Levin and Yehudai also identified source code changes that tend to lead to test changes, e.g., removing a class or changing a method signature~\cite{levin2017co}.
%Marsavina et al. used association rule mining to identify patterns of co-evolution~\cite{marsavina2014studying}. Their findings were later confirmed by Vidács and Pinzger~\cite{vidacs2018co}.
Such triggers have been used in past research to identify tests that need maintenance, based on supervised learning models~\cite{wang2021understanding} or by combining supervised learning with additional data sources, such as static analysis~\cite{liu2023drift} or code complexity metrics~\cite{huang2024towards}.

%Wang et al. also identified fine-grain source code change patterns---e.g., changes to \texttt{if-statements}---that may trigger test maintenance, and used these patterns to train a model to predict whether a test case is outdated~\cite{wang2021understanding}. Liu et al. improved the accuracy of this approach with more complex method of establishing traceability between source and test code and by automatically assigning code change patterns through static analysis, rather than manually labeling change patterns~\cite{liu2023drift}. Huang et al. also proposed a machine learning approach that predicts whether test code needs to be updated based on semantic inference from the code and code complexity metrics~\cite{huang2024towards}. 

Researchers have also developed approaches to automatically repair tests following source code changes.
Mirzaaghaei et al.'s approach is based on changes to method parameters and return values~\cite{mirzaaghaei2010automatically,mirzaaghaei2011automatic,mirzaaghaei2014automatic}, while Hu et al. infer edit patterns from datasets of source and test code changes~\cite{hu2023identify}.

LLMs have recently been applied to either identification or repair of obsolete tests. Hu et al. have applied a transformer model to identify and update obsolete tests~\cite{hu2023identify}. They do not separate these tasks into individual predictions, but perform a single repair task. If the final test is identical to the input, they assume that the test has been determined to still be relevant. Liu et al. use LLMs to update tests, but do not consider the identification task~\cite{liu2024fix}. Finally, Chi et al. have recently applied an LLM agent, using RAG, to identify and repair tests~\cite{chi2024reaccept}.

LLM agents have recently gained interest more broadly in software testing. Feldt et al. developed a taxonomy of agents and outlined an example conversational testing agent~\cite{feldt2023towards}. Yoon et al. have also demonstrated how LLM agents can perform Android GUI testing~\cite{yoon2023autonomous}. %Hong et al. also proposed a framework based on standard operating procedures to facilitate cooperation between agents and humans~\cite{hong2023metagpt}. As an example of how humans and LLM agents can interact, they model the roles in a software development team. 
Rasheed et al. have shown that cooperating LLM agents---each working on a particular subtask (e.g., code design, review, testing)---can generate higher quality code than a single LLM or LLM agent working alone~\cite{rasheed2024codepori}. %{\color{blue}Liu et al. uses static collector, neural reranker to help LLM model generate more correct test cases with more context~\cite{liu2024fix}. Chi et al. used dynamic validation to help LLM identify and change test cases.~\cite{chi2024reaccept}}

Our study is influenced by and expands on prior work. In terms of our exploratory contributions, we expand the range of triggers considered beyond past research, and we broadly explore the theoretical applications of automation in test maintenance---extending beyond prediction and update of individual obsolete test cases. In terms of our practical contributions, we offer a multi-agent architecture for prediction of obsolete test cases. This is similar to Chi et al.'s agentic approach~\cite{chi2024reaccept}. However, our approach splits the task into sub-tasks performed by multiple agents. We also have performed the first exploration and evaluation of LLM agents for test maintenance in an industrial setting and incorporate the experience and opinions of practitioners as an important source of data. 

%Shen et al. investigate the limitations of small LLMs when it comes to tool usage in LLM agents~\cite{shen2024small}.
%Their findings suggest that simplifying and dividing tasks into different instances can improve the performance of LLMs, especially LLMs of smaller sizes.

%Jiang et al. investigate the use of planning with LLMs, where the LLM first makes a plan for its actions before proceeding with those actions~\cite{jiang2023self}.
%Their findings indicate that a planning phase can improve performance, despite planning being an emergent ability of LLMs.
%Although this is not specifically about LLM agents it serves as support for one of the bases of LLM agents.
%Zhao et al. present a method of choosing between chain-of-thought and program-aided language models~\cite{zhao2023automatic}.
%Their findings indicate a benefit to performance for choosing the better-suited model for each problem.
%This is not specifically about LLM agents but also serves as a base idea for multi-agent frameworks, where multiple LLM agents cooperate and use their various specialised skills to collectively produce an improved result.

\section{Methods}\label{sec:methods}
In this study, we are interested in exploring the applications that LLMs or LLM agents could have within the test maintenance process. Specifically, we address the following research questions:
\begin{description}
    \item[RQ1] When changes occur in a project, what factors (e.g., changes to specific code elements) trigger the potential need for maintenance of test cases?
    \item[RQ2] What applications could LLMs or LLM agents have in the test maintenance process? 
    \begin{description}
        \item[RQ2.1] Which of the triggers from RQ1 could be acted upon by an LLM or LLM agent?
        \item[RQ2.2] What viable test maintenance actions could an LLM or LLM agent perform, based on these triggers?
        \item[RQ2.3] What considerations must be taken when deploying an LLM within an industrial environment?
    \end{description}
    \item[RQ3] What is the performance of our prototype multi-agent framework when predicting the need for test maintenance? 
\end{description}

\begin{figure}[!t]
    \centering
    \includegraphics[width=\textwidth]{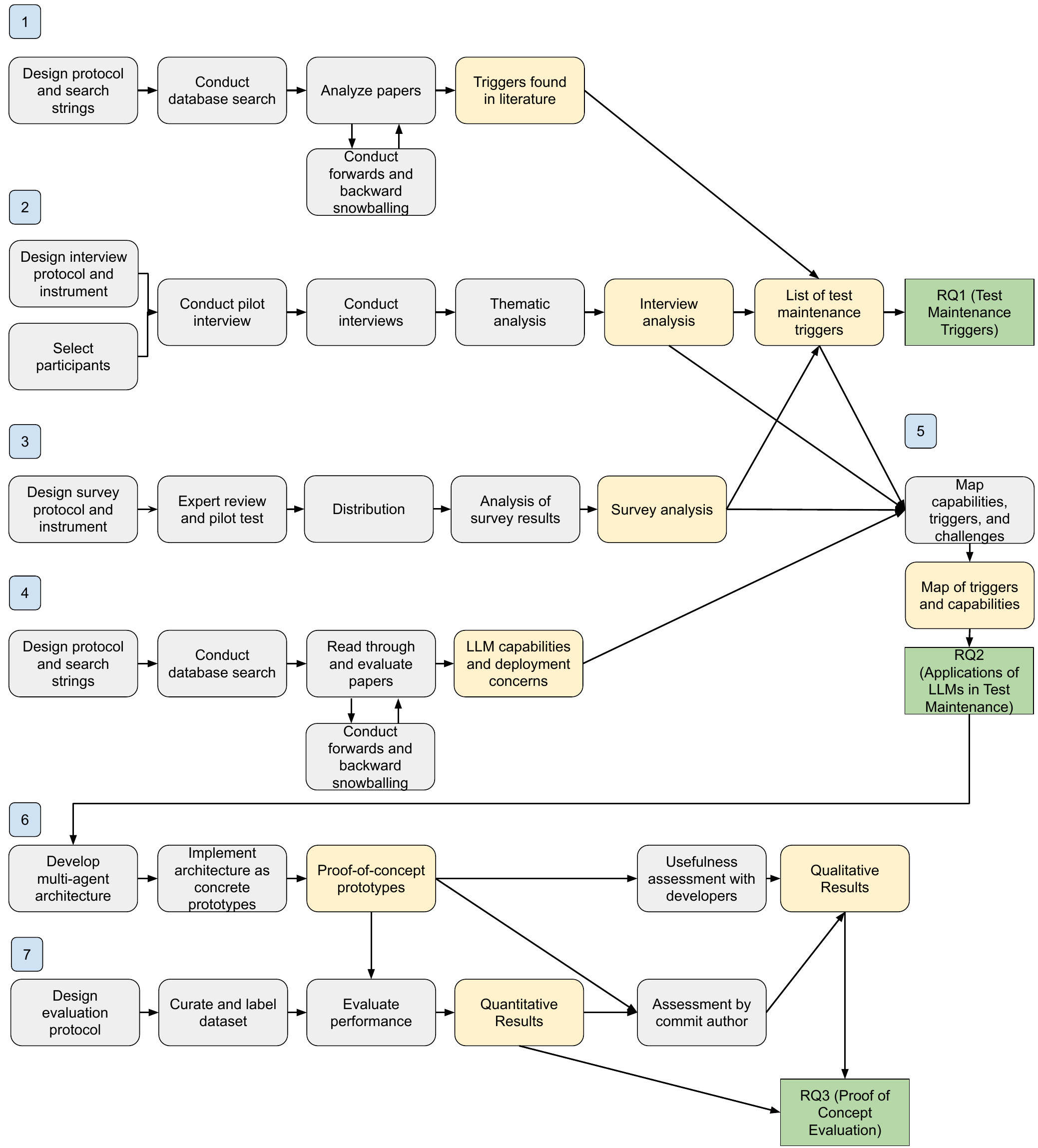}
    \caption{Overview of the case study. The numbers correspond to the steps outlined in Section~\ref{sec:methods}. Grey = activity, yellow = artifact, green = research question.} 
    \label{fig:outline}
    \vspace{15pt}
\end{figure}

The intention of \textbf{RQ1--2} is to offer an exploratory contribution to the field, investigating data signals that could indicate the need for test maintenance and to explore how these triggers or other sources of data could be used by LLMs or LLM agents. We then investigated considerations to be made when deploying LLMs in an industrial setting.

Towards a practical contribution, we implemented two prototypes, based on a multi-agent architecture that we propose for the purpose of predicting which test cases need to be updated following a change to the source code. The purpose of \textbf{RQ3} is to evaluate the  performance of these prototypes, including both a quantitative and qualitative analysis of their performance.

We address these research questions through a case study at Ericsson AB, a Swedish telecommunications company. Our case study follows the guidelines from Runeson and H\"{o}st~\cite{runeson2009guidelines}. To address the research questions, we performed the following steps (shown in Figure~\ref{fig:outline}):
\begin{enumerate}
    \item We performed a \textbf{literature review to identify test maintenance triggers} (Section~\ref{sec:rq1_review_method}). Data extracted from the identified publications partially address \textbf{RQ1}.
    \item We conducted \textbf{interviews} to understand the test maintenance process, triggers, and challenges at Ericsson, as well as potential tool support (Section~\ref{sec:rq1_rq2_interview_method}). Thematic analysis contributed to addressing \textbf{RQ1--2}. 
    \item To gather additional data, we conducted a \textbf{survey} at Ericsson regarding the test maintenance process, triggers, and challenges, as well as opinions regarding LLMs and tool support (Section~\ref{sec:rq1_rq2_survey_method}). Analysis of responses contributed to addressing \textbf{RQ1--2}.
    \item We performed a \textbf{literature review on LLM and LLM agent capabilities} to understand how LLMs have been used for software testing tasks, as well as to understand their suitability and limitations in test maintenance and industrial deployment (Section~\ref{sec:rq2_lit_method}). The results of this review contributed to addressing \textbf{RQ2}.
    \item We \textbf{mapped the capabilities of LLMs and LLM agents with test maintenance triggers and actions} identified in previous steps (Section~\ref{sec:rq2_mapping_method}). This process contributed to addressing \textbf{RQ2}. 
    \item We propose a \textbf{multi-LLM architecture}---and implemented two prototypes based on this architecture---to explore the viability of using LLMs to predict the need for test maintenance (Section~\ref{sec:rq3_design_method}). These prototypes are used to address \textbf{RQ3}. 
    \item We \textbf{evaluated the performance of the prototypes} using traditional machine learning measurements as well as qualitative analyses with Ericsson developers (Section~\ref{sec:rq3_eval_method}). This evaluation addresses \textbf{RQ3}. 
\end{enumerate}

\subsection{Literature Review to Identify Test Maintenance Triggers (RQ1)}
\label{sec:rq1_review_method}

We conducted a literature review to identify changes to project artifacts (e.g., to code, requirements, or other artifacts) or to the development process that may trigger a need for test maintenance. Our review was inspired by Keele's guidelines for systematic literature reviews~\cite{keele2007guidelines}. We applied the following search strings to ACM, IEEE, Science Direct, and Scopus:
\begin{itemize}
    \item (``test case'' OR ``test suite'') AND (update OR create OR refactor OR generate OR maintenance OR evolution OR management OR repair OR co-evolution) AND (``source code'' OR codebase) AND (factors OR criteria)  
    \item (``test case'' OR ``test suite'') AND (update OR create OR refactor OR generate OR maintenance OR evolution OR management OR repair OR co-evolution) AND ( ``foundational model'' OR ``machine learning'' OR ``LLM'' OR ``large language model'' )
\end{itemize}
The first was developed iteratively, with the intent of capturing relevant research on test maintenance while filtering publications that do not discuss potential triggers. The second adds a filter for LLMs, and was applied to ensure that no related work was missed. The strings were adapted for each database. In Scopus, the subject area was limited to computer science.

Database results were sorted by relevancy, then we selected publications by inspecting the title, followed by the abstract and conclusion. Inspection of results from each database ended after no new relevant papers were identified from three pages of results. We applied the following inclusion criteria:
\begin{itemize}
    \item Publications must have been published within the past 15 years---2009-2024---to ensure relevance to modern software development. 
    \item Publications must have been written in English. 
    \item Peer-reviewed and pre-print articles were considered. However, preference was given to peer-reviewed articles. 
    \item Publications must relate to test maintenance. 
    \item Publications must discuss factors that could trigger test maintenance. 
\end{itemize}

48 publications were found  to potentially meet all criteria. They were then read in full. Eight publications discussed test maintenance triggers and were retained for analysis. We performed backward and forward snowballing on these, yielding 64 additional potentially-relevant publications. From these, four additional publications were retained. 

We extracted triggers from these 12 publications. Then, we visually clustered them based on commonalities, such as belonging to particular levels of granularity within the code. We also merged redundant triggers. %We discuss the analysis of the identified publications and its results further in Section~\ref{sec:rq1_review_results}.

\subsection{Interviews (RQ1--2)}\label{sec:rq1_rq2_interview_method}
   
Interviews were held with Ericsson employees to better understand their test maintenance process, triggers, and challenges. The interviews also included questions about the potential use of LLMs and tool support. 

\begin{table}[!t]
    \centering
    \scriptsize
    \caption{Demographics of the interviewees. Experience refers to years of experience with testing, testing level refers to the level of granularity at which they regularly perform testing. Grouped participants were interviewed at the same time.}
    \begin{tabular}{|l|l|l|l|}
    \hline
       \textbf{ID} & \textbf{Experience (Years)} & \textbf{Role} & \textbf{Testing Level} \\ \hline \hline
       P1 & 15.00 & Developer & Unit \\ \hline %bruce
       \hline
       P2 & 3.50 & Data Scientist & Unit \\ \hline %carl
       \hline
       P3 & 6.00 & Developer & Unit \\ \hline %ingvr
       P4 & 5.00 & Developer & Unit \\ \hline %sreemanth
       P5 & 3.00 & Developer & Unit \\ \hline %tsigabu
       \hline
       P6 & 2.00 & Test Manager & Integration, System\\ \hline %anita
       P7 & 2.00 & Test Manager & Integration, System\\ \hline %hui
       \hline
       P8 & 25.00 & Principal Developer & Overseeing Process \\ \hline %ann-charlotte
    \end{tabular}
    \label{tab:demo_int}
\end{table}

\smallskip\noindent\textbf{Participant Selection:} We targeted Ericsson employees with experience in testing. We utilized convenience sampling, with elements of purposive sampling, to identify participants. Based on our knowledge of the organization, relevant teams were identified and invited to participate. Ultimately, we conducted five interviews, with eight participants. The interviews were conducted in the same manner regardless of the number of participants present. 

Table~\ref{tab:demo_int} presents the participant demographics. The most common role was ``developer'', followed by ``test manager''. The developers tested at the unit level, while test managers focused on integration and system level. The participants had a median of 4.25 (average of 7.69) years of experience. 

\begin{table}[!t]
    \centering
    \scriptsize
    \caption{Interview instrument.}
    \begin{tabular}{|l|p{12.2cm}|}
    \hline
    \multicolumn{2}{|c|}{\textbf{Demographic Questions}} \\ \hline 
    1 & What is your role?  \\ \hline
    2 & How many years of experience do you have with testing? \\ \hline
    3 & At what level do you perform testing most often? \\ \hline
    4 & Which programming language do you primarily work with? \\  \hline
    5 & How much time do you spend working on source code versus test code? \\ \hline
    6 & What is the proportion between test code and source code in your project? \\ \hline
    \multicolumn{2}{|c|}{\textbf{Test Maintenance Questions}} \\ \hline 
    7 & What reasons have you encountered for making changes to the test suite or a test case? \\ \hline
    & Can you give examples of the smallest code change that led to the largest test changes, the most common types of changes, the most tedious changes, or other unique cases? \\
    & What information do you use to make appropriate changes to the tests? \\
    & What kind of modifications do you usually apply to the tests? \\
    & What are factors do you consider when making changes? \\
    & How do changes in source code reflect changes in test code? \\ \hline
    8 & What specific changes in the source code lead to updates in the test suite? \\ \hline
    & Why does that source code change lead to a test suite change? \\  
    & Can you provide specific examples you have encountered? (E.g., adding new class or method or modifying an existing method, class, return statements, etc.) \\ \hline
    9 &  How often do you make changes to the test suite? \\ \hline
    & How often do you think the test suite needs to be changed? \\
    & Is this different from the actual update frequency? \\ \hline
    10 & Can you describe a normal sequence of events from a factor instigating a change in the test suite to when the change is complete? \\ \hline
    & What actions do you take as part of this process? \\ 
    & How much effort do the steps to update the test suite take? \\ 
    & Which step takes the most effort? \\ \hline
    11 & What consequences can there be from a change to the test suite? \\ \hline
    & Do these consequences affect how you approach your work? \\ 
    & How much additional maintenance effort can these consequences require? \\ \hline
    12 & What are the differences in the testing process between updating existing test cases versus creating new test cases? \\ \hline
    &  Are there different reasons or factors that cause each to occur? \\  
    & Are there generally differing amounts of effort required? \\ 
    & Which of these is the most common and why? \\ \hline
    \multicolumn{2}{|c|}{\textbf{Automation and LLM Questions}} \\ \hline
    13 & What tool assistance would you like with updating the test suite?  \\ \hline
    %& If you cannot think of anything: Suggestions on areas that need to be modified? Pseudo code for new or changed tests? Automatic IDE tools? \\ 
    & Which of these potential assistance use cases would be most valuable? \\ \hline
    14 & Do you currently use any tools to help update or create test cases? \\ \hline
    & Are any of these tools automated? \\
    & Do you think these tools, automated or not, work well? \\
    & If you have used LLMs, which, and how did you find the experience? \\
    & Would you like to use LLMs regularly as part of testing or test maintenance? \\
    & LLMs can potentially be used for creative purposes (e.g., providing advice) and boilerplate purposes (e.g., generating code). Do you have a preference for how you would use them? \\ \hline
    \end{tabular}
    \label{tab:instrument_int}
\end{table}

\smallskip\noindent\textbf{Interview Instrument:} We conducted semi-structured interviews (instrument in Table~\ref{tab:instrument_int}), so we could ask follow-up questions to gain further insight. Sessions were conducted online and were recorded for transcription. %All participants signed a consent form before beginning the interview. 
On average, each interview lasted 40 minutes. 

We performed a pilot interview, which led to removal of an irrelevant question and minor clarifications. Because changes were minor, the pilot interview was used in the final analysis.

\smallskip\noindent\textbf{Interview Analysis:} Interviews were first transcribed automatically, then manually corrected. We then performed thematic analysis following the guidelines described by Braun and Clarke~\cite{braun2006using}. 

We highlighted relevant parts of the transcripts and assigned ``code labels''---short identifiers---to each. We then developed ``codes'' describing each highlighted segment. Then, we iteratively grouped the codes and labels into themes and sub-themes. This process was completed by the second and third authors, with feedback from the others. We judged that we had reached saturation after analyzing the transcripts. %Therefore, we decided against conducting further interviews. %The results of this analysis are presented in Section~\ref{sec:rq1_rq2_interview_results}.

\begin{figure}[!t]
    \begin{subfigure}[t]{0.49\textwidth}
        \centering
        \includegraphics[width=\columnwidth]{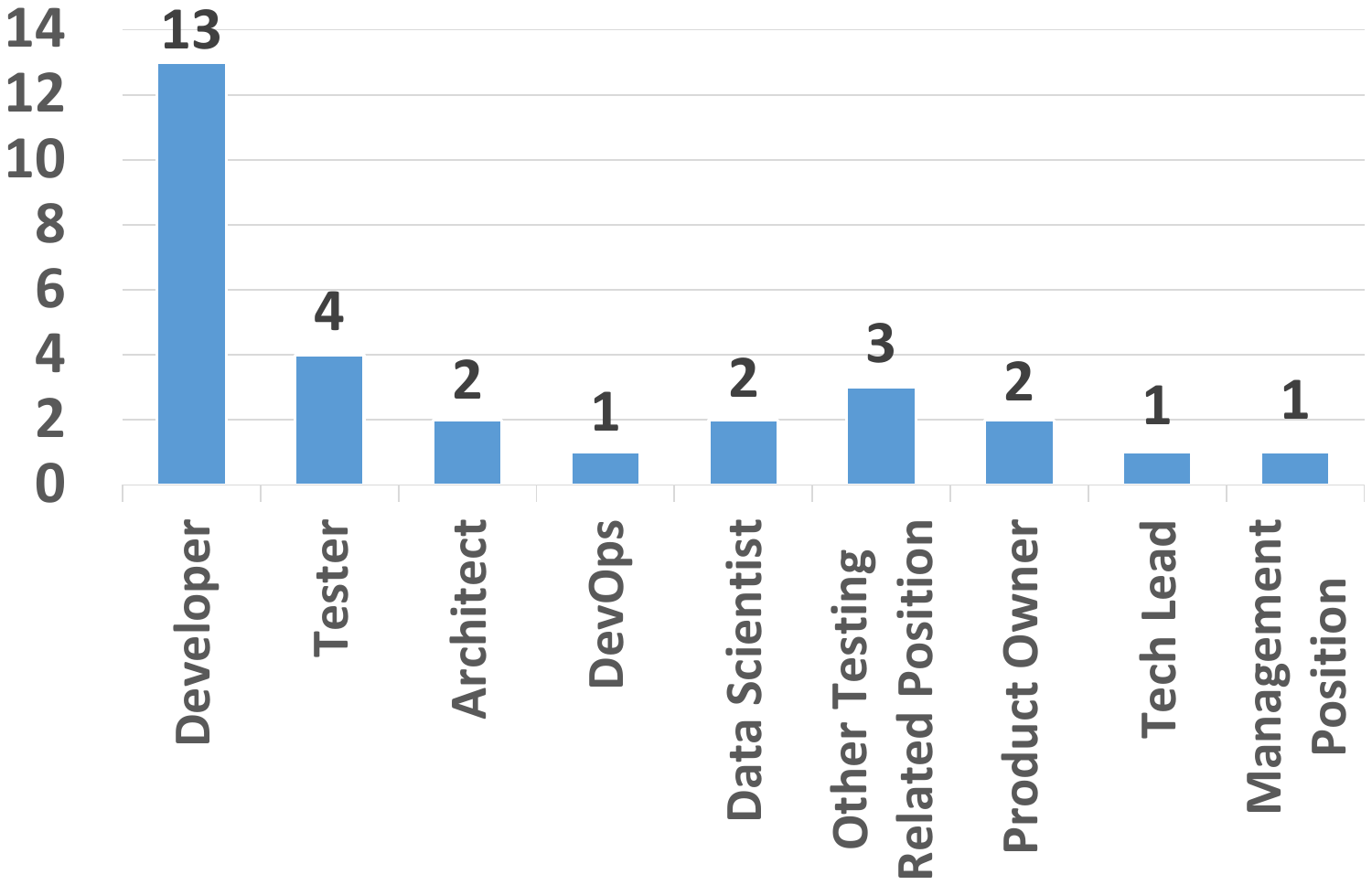} 
        \caption{Number of participants in each work role.}
    \end{subfigure} %
    \begin{subfigure}[t]{0.49\textwidth}
        \centering
        \includegraphics[width=\columnwidth]{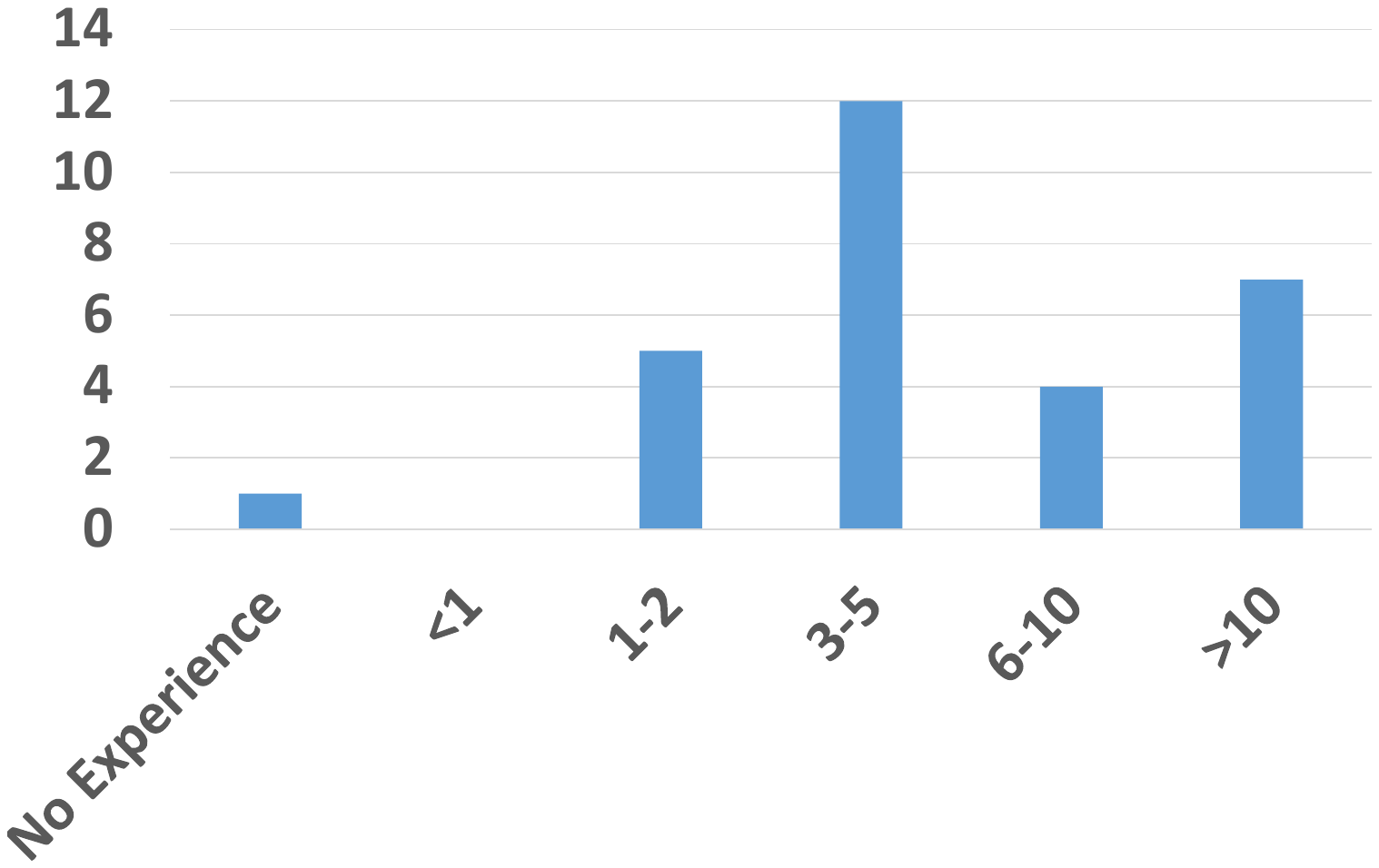} 
        \caption{Years of experience with software testing.}
    \end{subfigure} %
    \begin{subfigure}[t]{0.49\textwidth}
        \centering
        \includegraphics[width=\columnwidth]{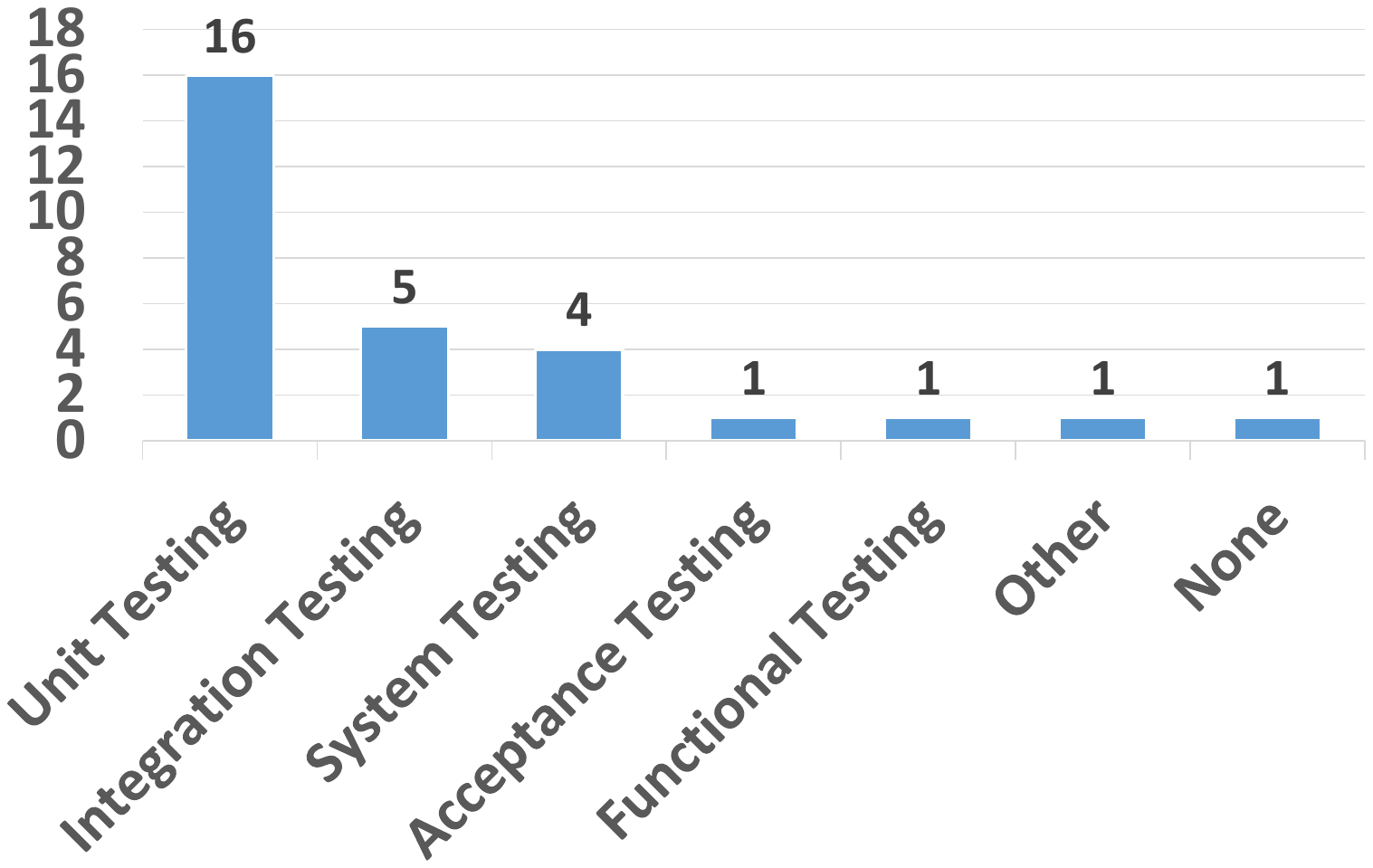} 
        \caption{Most common testing level.}
    \end{subfigure} %
    \begin{subfigure}[t]{0.49\textwidth}
        \centering
        \includegraphics[width=\columnwidth]{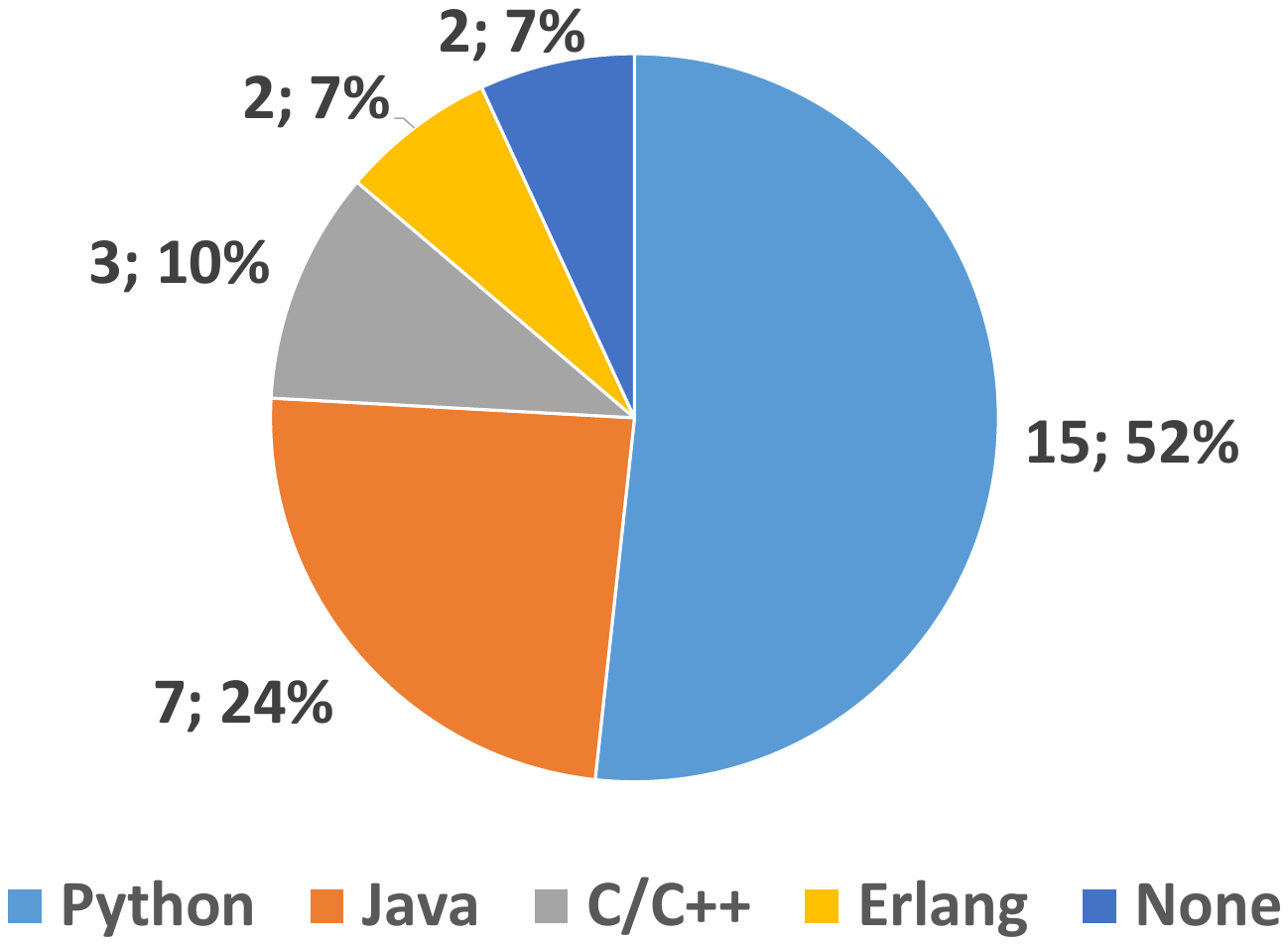} 
        \caption{Most common programming language.}
    \end{subfigure} %
\caption{Demographics of survey respondents.}
\label{fig:demo_survey}
\end{figure}

\subsection{Survey (RQ1--2)}\label{sec:rq1_rq2_survey_method}

As interviews require significant effort, we also conducted a survey to gain further insight into the test maintenance process, triggers, and challenges, as well as opinions regarding LLMs and tool support.

\begin{table}[!t]
    \centering
    \scriptsize
    \caption{Survey instrument.}
    \begin{tabular}{|l|p{4.9cm}|p{7.1cm}|}
    \hline
     & \textbf{Question} & \textbf{Response Options} \\ \hline 
     \multicolumn{3}{|c|}{\textbf{Demographic Questions}} \\ \hline 
     1 & What is your work role? & Developer, Tester, Architect, DevOps, Other  \\ \hline
     2 & How many years of experience do you have with testing? & None, $<$ 1, 1--2, 3--5, 6--10, $>$ 10 \\ \hline 
     3 & At what level do you perform testing most often? & Unit, Integration, System, Acceptance, Other \\ \hline
     4 & Which programming language do you primarily work with? & C/C++, Erlang, Go, Java, Julia, Python, Other  \\ \hline
    \multicolumn{3}{|c|}{\textbf{Test Maintenance Questions}} \\ \hline 
     5 & When making changes in the code, do you most often update source code or test code first? & Source, Test, Both Simultaneously, I Do Not Make Changes, Do Not Know \\ \hline 
     6 & What reasons have you encountered for making changes in the test suite or a test case? [max 4] & \begin{tabular}[p{5cm}]{@{}l@{}}Bug in Test Code, Bug in Source Code, \\Addition of New Feature, Changes in Requirements, \\Improve Test Coverage, Improve Test Performance, \\Improve Test Functionality, Changes in Tech Stack, \\Does Not Apply, Other \end{tabular} \\ \hline 
     7 & Which specific types of source code changes most commonly necessitate adjustments in the associated test code? [max 5] & \begin{tabular}[p{5cm}]{@{}l@{}}New Method, Remove Method, Change Method Body, \\Change Method Return Type or Value, Change \\Method Signature, New Class, Remove Class, Change \\Class Declaration, Change Object or Variable Type, \\Add Parameter, Remove Parameter, Change \\ Documentation, Change Loop Handling, Change \\Conditional  Statements, Change Single Line of Code \\ (e.g., assignment), Change Error or Exception \\ Handling, Change Interface or Class Hierarchy, Change \\ Interface, Change Parallelism/Concurrency, Other  \end{tabular} \\ \hline 
     8 & During the test maintenance process which step is most effort intensive? &  Identifying Solution, Designing Solution, Implementing Solution, Verifying Correctness of Solution,  Other \\ \hline 
    \multicolumn{3}{|c|}{\textbf{Automation and LLM Questions}} \\ \hline 
    9 & Have you used LLMs? & Yes--Often, Yes--Sometimes, Yes--Once or Twice, No--But Curious, No--Do Not Care, No--Do Not Want To, Do Not Know \\ \hline
    10 & How would you use LLMs during test maintenance? [max 3] & \begin{tabular}[p{5cm}]{@{}l@{}}Would Not Use, Code Generation, Improving My Code, \\Providing Suggestions, Suggesting Best Practices, \\Tracking Task Progress, Documenting Code, \\Understanding Code, Do Not Know,  Other  \end{tabular} \\ \hline 
    \end{tabular}
    \label{tab:instrument_survey}
\end{table}

\smallskip\noindent\textbf{Participant Selection:} Our audience was again Ericsson employees with testing experience, targeted using convenience sampling with purposive elements. We initially distributed the survey to a developer community, consisting of over 100 developers across different countries and sections within the company that work within the same product area. Due to a low initial response rate, we then distributed the survey to additional communities. The survey was open for 54 days. 

29 participants completed the survey. Their demographics of are shown in Figure~\ref{fig:demo_survey}. The most common role was developer (45\%), followed by testing-related positions (24\%). Participants had a median of 3-5 years of testing experience, and tested most often at the unit level. The most common programming language in use was Python. These demographic details are, broadly, similar to those of the interview participants. Therefore, we hypothesize that the interview and survey responses are complementary. 

\smallskip\noindent\textbf{Survey Instrument:} The survey instrument (Table~\ref{tab:instrument_survey}) was designed based on guidelines from Ghazi et al.~\cite{ghazi2018survey} and Kasunic~\cite{kasunic2005designing}. It was designed to take 5--10 minutes to ensure a reasonable completion rate. The intention was to provide quantitative data to augment the qualitative interview. Thus, questions were multiple-choice, rather than open-ended. Respondents could provide multiple answers to some questions, but the number of options they could select was limited to force  prioritization. The survey was designed prior to  analysis of interview data. 

We evaluated question wording using criteria established by Kasunic~\cite{kasunic2005designing} that offer rules for phrasing and structure of questions to minimize misunderstandings. The survey instrument was evaluated by a data analytics expert at Ericsson. It was also pilot-tested by three Ericsson developers to ensure there were no ambiguities in the questions, as well as to ensure it could be completed in less than ten minutes.

\smallskip\noindent\textbf{Survey Analysis:} The results of the survey were analyzed using descriptive statistics, and were contextualized by the thematic analysis of the qualitative data obtained during the interviews. %The results of the survey analysis are presented in Section~\ref{sec:rq1_rq2_survey_results}.

\subsection{Literature Review on LLM Capabilities for Test Maintenance (RQ2)}\label{sec:rq2_lit_method}

To understand how LLMs could potentially assist with test maintenance and considerations for industrial deployment, we conducted a literature review focusing on how LLMs have been applied within software testing. This review followed a similar procedure to the review in Section~\ref{sec:rq1_review_method}. We applied the following search strings to ACM, IEEE, Science Direct, and Scopus:
\begin{itemize}
    \item (llm OR ``large language model'' OR ``generative ai'') AND (``test management'' OR ``test maintenance'' OR ``software testing'') 
    \item (llm OR ``large language model'' OR ``generative AI'')  AND (``test case'' OR ``test suite'') AND (``software'')
    \item (llm OR ``large language model'' OR ``generative AI'') AND ( ``software engineering'' OR code) AND (suitability OR appropriateness OR abilities OR methods)
    \item (llm OR ``large language model'' OR ``generative AI'') AND ( trigger OR ``trigger point'' OR action OR apply OR automate OR improve OR simplify OR update OR updating OR create OR creating OR develop OR enable OR interact OR understand OR analyze OR generate OR generation) AND (``test management'' OR ``test maintenance'' OR ``test suite'' OR ``test case'' OR test OR ``test code'' OR ``quality assurance'' OR ``software testing'' OR ``software documentation'' OR ``test scenario'' OR ``test design'' ) AND (suitability OR appropriateness OR abilities OR methods)
\end{itemize}
These strings were iteratively developed. The first two capture publications where LLMs are used in software testing. The third to LLMs performing tasks involving source code. The fourth expands the  applied synonyms to capture additional publications. 

We applied the same inspection and stopping criteria previously discussed, with the following inclusion criteria:
\begin{itemize}
    \item Publications must have been published between 2017---when the first significant LLM-related developments emerged---and 2024.
    \item Publications must have been written in English. 
    \item Peer-reviewed and pre-print articles were considered. However, preference was given to peer-reviewed articles. 
    \item Publications must discuss the use of LLMs to perform tasks based on source or test code. The tasks performed must have some relevance to test maintenance. Publications must discuss the suitability (e.g., strengths and/or limitations) of LLMs to perform such a task.
    \item Alternatively, a publication must discuss considerations for industrial deployment of an LLM. 
\end{itemize}

We identified 67 publications for further examination. From this set, 10 were relevant for addressing RQ2. We performed forward and backward snowballing on these, which added an additional 92 publications. From those, two were deemed relevant, yielding 12 publications in total. %In total, we identified 12 publications with particular relevance for illustrating how LLMs could be used to automate actions related to test maintenance and considerations for industrial deployment. %We discuss these publications in more detail in Section~\ref{sec:rq2_lit_results}.

\subsection{Mapping Maintenance Triggers and Tasks with LLM Capabilities (RQ2)}\label{sec:rq2_mapping_method}

To identify how LLMs or LLM agents could assist with test maintenance, we compared  data gathered from the literature review on test maintenance triggers, the interviews, and the survey. We mapped this data to the capabilities of LLMs demonstrated in previous literature. 

To decide which specific maintenance tasks LLMs or agents could perform, we utilized a mind-mapping process. This was done by brainstorming which LLM actions and applications might fit specific triggers, and how these could be combined to solve test maintenance tasks and challenges identified in the interviews and survey. We also considered how LLMs should be implemented to match developers' preferences and ways of working.

%As part of addressing RQ1, we split test maintenance triggers into high- and low-level triggers, where high-level triggers relate to process-level decisions (e.g., requirement changes or a decision to improve the efficiency of the test suite) and low-level triggers referring to individual changes to project artifacts. We created one map for high-level and one for low-level triggers. 

\subsection{Proof-of-Concept Design (RQ3)}\label{sec:rq3_design_method}

To explore the practical applicability of current-generation LLM agents, we have developed a proof-of-concept multi-agent framework that observes changes to the source code, then predicts which test cases require maintenance. This section presents the architecture for this framework, as well as details on two concrete implementations of this architecture.
%We iteratively explored multiple potential architectures for a framework that could predict the need for test maintenance. Ultimately, we implemented two prototypes in Python, based on one proposed architectures. 
%This section presents the architecture and implementation details.

\smallskip\noindent\textbf{Overall Architectures:} We focus on an agent-based architecture, where retrieval-augmented generation is used to obtain information from the source and test code, and where memory capabilities are incorporated to enable  complex multi-step processes and correction of previous mistakes. %LLM agents typically demonstrate improved performance over non-agent LLMs, especially in zero-shot prompting interactions~\cite{wang2024survey, xi2023rise}. %We focused on multi-agent models of collaboration where individual agents or LLM instances completed their own assigned aspects of a test maintenance task.  
%An alternative to an LLM agent would be to pre-train an LLM for a specific task. However, as the costs of pre-training can be very expensive, we decided that an LLM agent would be more appropriate. In addition, pre-training is not as adaptive as an agent architecture as the model would still need to be retrained to gain access to updated information. However, these two concepts are complementary---a pre-trained model could be used within an agent architecture---and their combination could be explored in future work. 
%We developed one core architecture for performing the prediction tasks, which employs a multi-LLM architecture to improve performance~\cite{li2024more, hong2023metagpt, guo2024large}. This allows 
Each LLM agent and individual LLM instance in the architecture focuses on a particular subtask, and each agent can be developed to employ specialized tools. 

\begin{figure}[!t]
    \centering
    \includegraphics[width=0.8\textwidth]{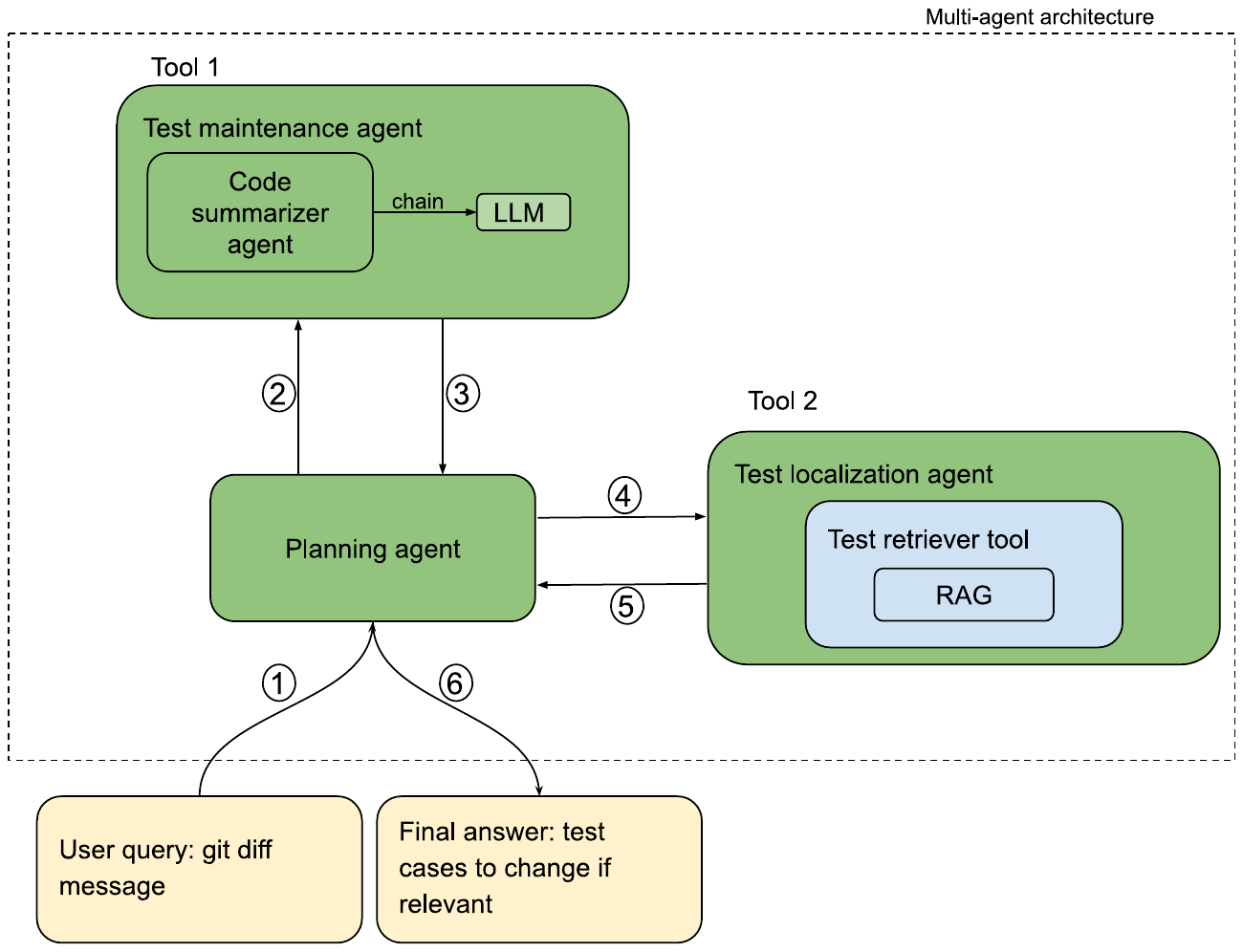} \vspace{-40pt}
    \caption{Architecture incorporating multiple LLM agents (dark green). Yellow represents input and output, light green for LLM instances, and blue for tools.}
\label{fig:architecture_planning}
\end{figure}

This architecture (shown in Figure~\ref{fig:architecture_planning}) employs a planning agent that coordinates the effort of other LLM agents,  following this sequence of steps: 
\begin{enumerate}
    \item A code change (\texttt{git diff}) is provided to the framework.
    \item The planning agent provides the code change to the test maintenance agent to determine whether test maintenance is necessary.
    \begin{itemize}
        \item The test maintenance agent requests a summary of the change from the  code summarizer agent.
        \item The test maintenance agent will provide the summary to an LLM instance to determine whether test maintenance is necessary.
    \end{itemize}
    \item The test maintenance agent sends the result to the planning agent. If  maintenance is necessary, the agent will send a confirmation and the change summary. Otherwise, the agent will send an explanation, which the planning agent will use to provide a final response.
    \item When test maintenance is  necessary, the planning agent will contact the test localization agent, sending the code change and the summary from the test maintenance agent.
    \begin{enumerate}
        \item The test localization agent invokes the test retriever tool, which uses RAG to retrieve relevant test cases based on the input.
        \item The retriever tool will either fetch test code or summaries of test code, depending on which prototype is being used.
    \end{enumerate}
    \item The test localization agent responds to the planning agent with a list of test cases that need to be changed.
    \item The planning agent responds to the user, either indicating that  test maintenance is unnecessary or providing information on which test cases need to be changed.
\end{enumerate}

% \begin{figure}[!t]
%     \centering
%     \includegraphics[width=0.8\textwidth]{figure/multi-agent_chain.pdf}
%     \caption{Pipeline-based architecture employing test case summaries. Yellow represents input and output, light green for LLM instances, dark green for LLM agents, and blue for tools used by the agents.}
%     \label{fig:architecture_chain}
%     \Description[Architecture of the LLM chain architecture without a planner agent.]{Architecture of the LLM chain architecture without a planner agent, the architecture is described in the text below.} 
% \end{figure}

% The pipeline-based architecture (Figure \ref{fig:architecture_chain}) makes predictions through the following steps:
% \begin{enumerate}
%     \item The user queries the chain with the changed source code.
%     \item The RAG tool retrieves the current source code and sends it to the code summarizer. The code summarizer also receives the user query---the new code---for comparison.
%     \item The summarizer sends a summary of the code change to the test maintenance trigger comparator. 
%     \item The trigger agent determines whether test maintenance is necessary and, if so, also includes the name of the relevant code method.
%     \item If test maintenance is necessary, the test localizer retrieves test case summaries. 
%     \item The test localizer outputs the list of relevant test cases to the user or states that test maintenance is unnecessary.
% \end{enumerate}

\smallskip\noindent\textbf{LLM Agent and Instance Implementation:} We have implemented two versions of this architecture. The first examines the test code directly to determine whether a test requires maintenance. The second employs an LLM to generate a natural language summary of all test cases beforehand, then acts on that summary instead of the code. 

The agents are implemented using the LangChain framework~\cite{langchain}. %This framework was chosen because it offered implementations of functionality including agent wrapper methods, vector stores, embedding model, prompt engineering and tool-calling. This allowed us to focus on implementing our particular use case rather than reinventing core aspects common to other LLM agents.
The LLM agents and instances employ the Qwen2.5-32B-Instruct model~\cite{qwen2.5}. Qwen2.5-32B-Instruct is a version of the Qwen2.5-32B model that has been fine-tuned to follow instructions. %DeepSeek-R1-Distill-Qwen-14B model~\cite{deepseekai2025deepseekr1incentivizingreasoningcapability} is also used during the development phase for tweaking the prompts for the agent, as it prints out its entire thought process while analyzing queries. This helps identify issues when the agent does not behave as expected based on the prompts, by providing clues through the model’s output. However, eepSeek-R1-Distill-Qwen-14B was eventually discarded due to its frequent hallucinations. 
Ericsson must approve LLMs for internal use. %\textcolor{red}{Nasser suggesting an alternative for "At the time of the implementation, we determined that Mistral 7B - Instruct was the best of the approved models for our use case." in what follows. If ok, just remove the red text.} 
At the time of implementation, we determined that Qwen2.5-32B-Instruct was the best model for our use case, considering availability, capabilities, and the Ericsson approval process and guidelines. 

\begin{figure}[!t]
    \centering
    \includegraphics[width=0.6\textwidth]{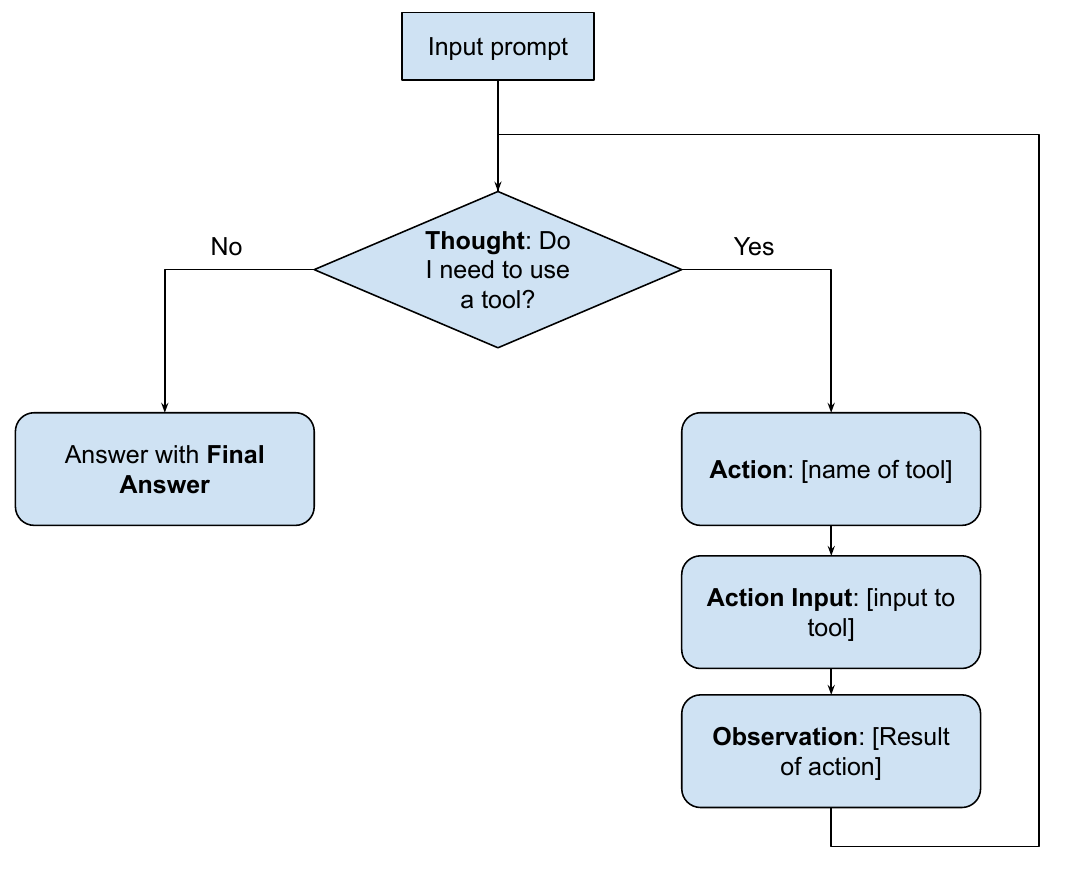}
    \caption{Flow diagram showing how a ReAct agent invokes tools to perform tasks.}
    \label{fig:react_cycle}
\end{figure}

The LLM agents are implemented following the ReAct agent structure~\cite{yao2022react}, visualized in Figure~\ref{fig:react_cycle}. After receiving a prompt, the model goes through a cycle of Thought-Action-Observation, where an action is taken based on a thought, and an observation is made from the results of the action. 

If the observation is satisfactory, the agent stops, and if not, it continues to iterate until an answer or a timeout is reached (maximum 3 iterations for each LLM agent in our implementation). If the agent fails to answer using the correct format, it will be reminded what the correct format looks like and asked to answer again (counting as an iteration). We also impose a limit of 300 seconds per prompt. 

Each agent is implemented with its own individual memory, planning mechanisms, and tools. The memory is also printed to \texttt{stdout} to provide additional context to explain an agent's decisions. Each agent has an individual prompt that gives them instructions on their role and how to act. 

LLM instance are invoked through a chain of prompts. As they are not full agents, they do not have access to memory or tools. However, they offer a faster response. Tools are utilized through separate calls that then send results as input to the LLM instances. The prompts provided to each LLM agent or instance are provided in our supplementary data package\footnote{Available at \url{https://doi.org/10.5281/zenodo.13341060}}.

The planning agent controls the overall process.  It invokes other agents as tools that it can utilize. If it determines that their output is insufficient, it can invoke them again. Its prompt specifies that it needs to use these ``tools'', expands on its tasks, specifies the input format (\texttt{git diff}), and offers additional instructions. %After the agent has received a query, it can choose how to approach the problem, which tools to use, and when it considers the problem solved. These decisions are made by the model. 

The test maintenance agent invokes the code summarizer agent as a tool to construct a natural language summary of the \texttt{git diff}. It then provides that summary to a simple LLM instance to determine whether test maintenance is likely to be necessary. 

%LLM consists of a cooperating LLM agent and instance, as can be seen in Figure~\ref{fig:architecture_planning}. The summarizer is implemented as an agent with potential tool access for scalability, while a simple LLM instance was used determining if test maintenance is needed or not based on the change summary. The test maintenance agent is then guided by its prompt to make decision if test maintenance is necessary or not.

% For the architecture without the planning agent, the code summarizer agent is instead implemented as an LLM instance. The RAG tool for the source code is called separately first from the user query and then the results from the RAG tool, along with the user query, are sent to the LLM to be inserted in the prompt instead of utilizing a tool structure. The agents and LLM instances are also connected in a chain where the outputs of the LLMs and agents serve as inputs for the following LLM or agent.

The test localization agent uses a RAG tool to access either the raw test code, or natural language summaries of test cases, for the specific commit that the \texttt{git diff} references. This tool is used to determine which test cases are relevant to the code change.
%has its own prompt and access to a test retriever tool which make use of retrieval-augmented generation (RAG) that includes an embedded vector store for the corresponding version of the relevant code repository. This vector store includes either the vectorized test codes or vectorized test codes summaries, depending on which version of the prototype being used, for the specific commit from which the git diff originates. For example, a new change in the latest code repository will reference the code at the tip of the main branch, where the corresponding test code or test code summaries are vectorized. %In the chain architecture, the test localisation agent works directly with output from the trigger comparator. It chooses on its own whether to search for test cases or not depending on the output of the trigger comparator.
The RAG tool uses the BAAI/bge-m3 embedding model~\cite{bge-m3} to encode code for processing by an LLM. This was the best embedding model in current deployment at Ericsson. %that we had access to for batch use. %We also tested other non-embedding models, but these offered worse performance than BAAI/bge-m3. 
We employed the Facebook AI Similarity Search (FAISS) library~\cite{douze2024faiss} as a vector store for the embedded vectors.

\subsection{Proof-of-Concept Evaluation (RQ3)}\label{sec:rq3_eval_method}

We evaluated the performance of the two prototypes. We then performed two validations with Ericsson developers---the first a general examination of the usefulness of the prototypes, and the second an assessment of the prototypes' predictions by a developer responsible for a subset of the commits used to benchmark the prototypes. Together, these evaluations shed light on the current capabilities and limitations of LLM agents to predict the need for test maintenance. 

\smallskip\noindent\textbf{Evaluation Dataset:} We developed a dataset from the commit history of a Java-based repository at Ericsson. The repository contains code for a a microservice that performs data processing tasks. The test code contains unit and integration tests.

Our ground truth is based on the assumption that source and test cases in the same commit represent co-evolution. The individual code changes are captured in the form of the \texttt{git diff} files for that commit. The following code changes were excluded:
\begin{itemize}
    \item Commits with no changes to the source code.
    \item All changes to test code that were not encapsulated within a test case, such as changes to test support code.
\end{itemize}

The dataset spans 54 commits, which collectively contain 634 distinct code changes. The default \texttt{git diff} only includes three context lines above and below each change. We extended this to 9 nine lines to incorporate more contextual information around the changed code. This number was selected based on informal experimentation.

\smallskip\noindent\textbf{Evaluation Procedure:} For each commit, we performed the following\footnote{The per-commit results are available in our supplementary data package: \url{https://doi.org/10.5281/zenodo.13341060}} : 
\begin{enumerate}
    \item The information stored in the RAG tool was updated to reflect the current commit.
    \item Each individual code change was input  into the agent setup. These output was saved. If maintenance was necessary, we extracted the names of suggested test cases.
    \item Ground truth is known for a commit, not for each code change. To assess accuracy, the suggested test cases for all code changes in that commit were compared to the test cases changed in the actual commit.
\end{enumerate}

In this experiment, we used a  temperature of 0 to limit stochasticity. However, we performed the evaluation twice, as there may still be some variance. As the results were very similar across both trials, we did not perform further repetitions. For each commit, we define:
\begin{itemize}
    \item A \textbf{true positive (TP)} is a test predicted as \textit{requiring} maintenance that \textit{actually} required maintenance.
    \item A \textbf{true negative (TN)} is a test predicted as \textit{not requiring} maintenance that \textit{actually} did not require maintenance.
    \item A \textbf{false positive (FP)} is a test predicted as \textit{requiring} maintenance that \textit{did not} require maintenance.
    \item A \textbf{false negative (FN)} is a test predicted as \textit{not requiring} maintenance that \textit{actually}  required maintenance.
\end{itemize}

The commits in the dataset reflect two distinct outcomes. In 32 commits, test maintenance was \textit{required}. In the remaining 22 commits, \textit{no test cases were updated}. We split these cases when evaluating performance for two reasons. First, in cases where no maintenance was required, there can be no TP or FN, only TN and FP. Second, the commits for each case are imbalanced in the number of code changes per commit. In cases where tests were updated, there is an average of 16.7 changes per commit. In contrast, the cases where maintenance was not required average only 4.5 changes per commit. 

\noindent We calculate the following metrics:
\begin{itemize}
    \item Accuracy = $\frac{TP+TN}{(TP+FP+TN+FN)}$
    \item Recall = $\frac{TP}{(TP+FN)}$
    \item Precision = $\frac{TP}{(TP+FP)}$
    \item F1 score = $2 * \frac{(precision * recall)}{(precision + recall)}$
    \item F2 score = $(1 + 2^2)*  \frac{(precision * recall)}{(2^2*precision + recall)}$
\end{itemize}

We use these measurements to provide multiple viewpoints on performance. In particular, as the datasets contain a large number of TN, the F1 and F2 scores are useful because they focus on TP, FP, and FN. While the F1 score is commonly used to assess machine learning tasks, we also include the F2 score, which places a heavier weight on recall than precision. We do this because we hypothesize that recall is more important than precision in this task---omitting tests that should have been included in a prediction is potentially more detrimental than including additional false positives.

For the subset of commits where test maintenance was required, we report all five metrics. For the subset of commits where test maintenance was not required, we only report accuracy, as TP and FN outcomes are not possible.

\smallskip\noindent\textbf{Usefulness Validation:} We held an informal validation of the prototypes with two developers from Ericsson. %One had participated in the interviews, while the other is one of the co-authors of this study (Roy Liu). Thus, 
Both were familiar with LLMs and the specific context of this study. During the validation, the developers were shown the prototypes, the results of the experiment, and examples of their output. They were asked to provide reflections on what adaptations or requirements would need to be met to deploy the prototypes in practice.

\begin{table}[!t]
    \centering 
    \scriptsize
    \caption{Interview with commit author.}
\label{tab:post_interview_table}
    \begin{tabular}{|l|p{12cm}|}
    \hline
    \multicolumn{2}{|c|}{\textbf{Session 1: Questions about test maintenance}} \\ \hline 
    1 & When you modify the code, how do you identify which test cases need to be updated?   \\ \hline
    & How difficult do you think it is to locate the correct test cases to update?\\
    & How confident are you that you have identified all relevant test cases?\\
    & If all tests pass for your change, do you still update any test cases?\\ \hline
    \multicolumn{2}{|c|}{\textbf{Session 2: Questions regarding FPs}} \\ \hline 
    2 & What is your impression of these FPs (\textbf{before} checking the prototype output)?  \\ \hline
    & Do the FPs offer any useful hints about which test cases might need to be changed? \\
    & Do you think any of them might actually be TPs?  \\ \hline
    3 & What is your impression of these FPs (\textbf{after} checking the prototype output)?  \\ \hline
    & When you think that output for a FP is \textbf{useful}, what do you mean? \\
    & When you think that output for a FP is \textbf{not useful}, what do you mean?  \\ \hline
    \multicolumn{2}{|c|}{\textbf{Session 3: Questions after checking FPs}} \\ \hline
    4 & Which metric matters more to you, recall or precision?  \\ \hline
    5 & If you are using this tool, how fast do you want it to respond to, e.g., one code change?  \\ \hline
    6 & For the FP that you feel should actually be TP, why do you think you missed those cases? \\ \hline
    7 & What other suggestions do you have regarding the prototype? \\ \hline
    \end{tabular}    
\end{table}

\smallskip\noindent\textbf{Assessment by Commit Author:} Finally, we conducted an interview with the primary developer responsible for most of the commits in the dataset. %During this interview, we presented three commits selected from the dataset and the output of the prototype based on test summaries for each of these commits, focusing on FP results. 
The interview questions are listed in Table~\ref{tab:post_interview_table}.

The interview was conducted in three sessions. In the first, we asked general questions about test maintenance. In the second, we presented three commits selected from the dataset, performance results for those commits, and a list of 38 test cases identified as FPs. These results were from the prototype that utilizes test summaries. 

We asked the developer to classify each suggestion as ``useful'' or ``not useful'', and to explain why. We then presented the prototype's output for each FP---the explanation of why the prototype thinks the test potentially needs to be updated---and asked the developer to re-examine their previous comments on each FP. The developer determined whether the output was useful or not. A suggestion was deemed potentially useful if:
\begin{itemize}
    \item The test could be affected by the changed code. 
    \item The explanation is substantiated.
    \item The suggestion is either correct, or if not correct, could benefit the test suite or  maintenance process (e.g., suggests new tests to create).
\end{itemize}

In the final session, we asked questions about their performance preferences when using the prototype as part of their workflow.

\section{Results}\label{sec:results}
In this section, we present data and observations from our research activities, which are synthesized into answers to the research questions in Section~\ref{sec:discussion}.

\subsection{Literature Review to Identify Test Maintenance Triggers (RQ1)}\label{sec:rq1_review_results}

We define the test maintenance triggers that we identified from literature in Tables~\ref{tab:triggers_func}--\ref{tab:triggers_line}. Broadly, the triggers can be divided \textit{documentation changes}---e.g., comments, user manuals, or requirements---and \textit{source code changes}. Code changes can also be grouped based on the level of granularity---functionality, class-level, method-level, and line-level \footnote{There is some overlap in triggers---e.g., ``addition of functionality'' could include ``addition of a class'' or ``addition of a method''. However, we retain ``functionality'' to capture cases where it was not clear what form the functionality took.}.

\begin{comment}
\begin{figure}[!t]
    \centering
    \includegraphics[width=\textwidth]{figure/test_factors_from_lit.pdf} 
    \caption{Test maintenance triggers identified in the literature review. The triggers are grouped into clusters, indicated by color.}
    \label{fig:triggers_overview} 
\end{figure}
\end{comment}

\begin{table}[!h]
    \centering
    \scriptsize
    \caption{Test maintenance triggers that affect the general functionality of the system.} 
    \begin{tabular}{|l|p{6cm}|l|}
    \hline
       \textbf{Type of Change}  &  \textbf{Description} & \textbf{Sources} \\ \hline 
       \textbf{Changes to Functionality}  & \multicolumn{2}{|c|}{\textbf{Code changes that affect system behavior.}}
     \\ \hline 
       Add Functionality  & Implementing new functionality. This is a general trigger for cases where it is not clear how the functionality was implemented. & \cite{shimmi2022patterns, levin2017co, mirzaaghaei2011automatic}\\ \hline
       Add Interface  & Implementing a new interface. & \cite{mirzaaghaei2012supporting, levin2017co, mirzaaghaei2014automatic, mirzaaghaei2011automatic}\\ \hline
       Error and Exception Handling & Changes to code that handles errors/exceptions, such as \texttt{throw}, \texttt{try}, and \texttt{catch} keywords. & \cite{liu2023drift}\\ \hline
       Parallelism and Concurrency  & Changes to how the system handles parallel threads and concurrency, such as the \texttt{synchronised} keyword.  & \cite{liu2023drift} \\ \hline 
       Remove Functionality  & Code is removed. This is a general trigger for cases where it is not clear how the functionality was removed. & \cite{shimmi2022patterns, levin2017co}\\ \hline 
       Update API  & Changes to how an external API is invoked. & \cite{hu2023identify}\\ \hline
    \end{tabular}
    \label{tab:triggers_func}  
\end{table}

\begin{table}[!h]
    \centering
    \scriptsize
    \caption{Test maintenance triggers that describe changes made to a class.} 
    \begin{tabular}{|l|p{6cm}|l|}
    \hline
       \textbf{Type of Change}  &  \textbf{Description} & \textbf{Sources} \\ \hline 
       \textbf{Changes to a Class}  &  \multicolumn{2}{|c|}{\textbf{This grouping covers changes made at the class-level.}}  \\ \hline
       Add Class & Implementing a new class. & \cite{marsavina2014studying, mirzaaghaei2012supporting, liu2023drift, reich2023testability, levin2017co, vidacs2018co}\\ \hline
       Class Declaration & Changes made to a class declaration, including extending class hierarchy. & \cite{marsavina2014studying, levin2017co, wang2021understanding, vidacs2018co, mirzaaghaei2014automatic, mirzaaghaei2011automatic}\\ \hline
       Constructor & Changes made to the class constructor, e.g., creating a constructor or adding a  parameter. & \cite{reich2023testability}\\ \hline
       Fields & Changes made to the fields of a class. & \cite{marsavina2014studying, wang2021understanding, vidacs2018co}\\ \hline
       Remove Class & Removing a class that previously existed. & \cite{marsavina2014studying, levin2017co, vidacs2018co}\\ \hline
    \end{tabular}
    \label{tab:triggers_class} 
\end{table}

\begin{table}[!h]
    \centering
    \scriptsize
    \caption{Test maintenance triggers that describe changes made to a method.} 
    \begin{tabular}{|l|p{6.5cm}|p{2.5cm}|}
    \hline
       \textbf{Type of Change}  &  \textbf{Description} & \textbf{Sources} \\ \hline 
       \textbf{Changes to a Method}  &  \multicolumn{2}{|c|}{\textbf{This grouping covers changes made at the method-level.}}  \\ \hline 
       Add Method & Implementation of a new method. This includes adding an overloaded or overridden method. & \cite{mirzaaghaei2012supporting, reich2023testability, vidacs2018co, mirzaaghaei2014automatic, mirzaaghaei2011automatic} \\ \hline
       Method Declaration & Changes to a method declaration or signature. This includes changing access level, adding or removing \texttt{final/static} keyword, changing value or type of the return, and adding, removing and changing a parameter type. & \cite{mirzaaghaei2012supporting, marsavina2014studying, liu2023drift, reich2023testability, levin2017co, wang2021understanding, vidacs2018co, mirzaaghaei2014automatic, mirzaaghaei2011automatic, mirzaaghaei2010automatically, liu2023drift, levin2017co, mirzaaghaei2014automatic, mirzaaghaei2010automatically} \\ \hline
    \end{tabular}
    \label{tab:triggers_method} 
\end{table}

\begin{table}[!h]
    \centering
    \scriptsize
    \caption{Test maintenance triggers that describe changes made to a single line of code.}
    \begin{tabular}{|l|p{7.5cm}|l|}
    \hline
       \textbf{Type of Change}  &  \textbf{Description} & \textbf{Sources} \\ \hline 
       \textbf{Changes to LOC}  & \multicolumn{2}{|c|}{\textbf{Changes made to the source code that affect a single line.}}  \\ \hline 
       Arrays  & Changes to arrays, e.g., new array or changing  access level. & \cite{liu2023drift}\\ \hline
       Assignments  & Changes to value assigned or compound assignments. & \cite{liu2023drift}\\ \hline 
       Attribute & Declaration of attributes and extraction for assertion. & \cite{marsavina2014studying, reich2023testability, vidacs2018co}\\ \hline
       Modifiers & Changing and updating modifiers, e.g.,  \texttt{public} keyword. & \cite{liu2023drift, hu2023identify}\\ \hline
       Identifier  & Changes to identifiers, e.g., variable or package names. & \cite{liu2023drift, hu2023identify, wang2021understanding}\\ \hline
       Object State & Removing or adding object state. & \cite{levin2017co}\\ \hline
       Override System Time & Overriding the system time. & \cite{reich2023testability}\\ \hline 
       Statement & Changes to control-flow, conditional, or loop statements  & \cite{marsavina2014studying, liu2023drift, levin2017co, wang2021understanding} \\ \hline
       Type & Modify data type, either keyword or by casting. & \cite{liu2023drift} \\ \hline
       Variables  & Changes to the declaration of a variable.  & \cite{liu2023drift}\\ \hline
    \end{tabular}
    \label{tab:triggers_line} 
\end{table}

\subsection{Thematic Analysis of Interviews (RQ1--2)}\label{sec:rq1_rq2_interview_results}

%In total, five major themes were identified during the analysis of the interview transcripts---\textit{Triggers}, \textit{Quality Assurance},  \textit{Challenges}, \textit{Desired Tool Support}, and \textit{Attitudes Towards LLMs}. Each has multiple sub-themes. 

An overview of the themes identified during analysis of the interviews is presented in Table~\ref{tab:interviews_overview}, and we discuss each theme and sub-theme below. We also indicate the number of codes corresponding to each theme or sub-theme, as well as the percentage of total codes (for all themes, or for all sub-themes of a particular theme)\footnote{A high number does not necessarily indicate higher importance, but does indicate that participants had more to say about the subject.}. 

%In addition, in Figure~\ref{fig:interview_quant}, we indicate the number of codes corresponding to each theme and sub-theme. The number of occurrences of a theme or sub-theme does not necessarily indicate its relative importance, but it does indicate that the interview participants had more to say about some subjects. 

\begin{table}[!t]
    \centering
    \scriptsize
    \caption{Overview of interview themes, with description and number of codes (and percentage of total) that correspond to that theme.}
    \begin{tabular}{|l|p{7.5cm}|l|}
    \hline
       \textbf{Theme}  &  \textbf{Description} & \textbf{Num (\%)}  \\ \hline 
       Triggers & Reasons for performing test maintenance. & 129 (25\%) \\ \hline
       Quality Assurance & Ways to ensure that test cases and suites hold high quality. & 212 (41\%)\\ \hline
       Challenges & Current issues experienced with test maintenance. & 76 (15\%) \\ \hline
       Desired Tool Support & Type of automated help wanted with test maintenance. & 69 (13\%) \\ \hline
       Attitudes Towards LLMs& Attitudes about using LLMs for test maintenance. & 28 (6\%)\\ \hline
    \end{tabular}
    \label{tab:interviews_overview}
\end{table}

\begin{comment}
\begin{figure}[!t]
    \begin{subfigure}[t]{0.49\textwidth}
        \centering
        \includegraphics[width=\columnwidth]{figure/themes/themes.pdf} 
        \caption{Overall Themes.}
    \end{subfigure} %
    \begin{subfigure}[t]{0.49\textwidth}
        \centering
        \includegraphics[width=\columnwidth]{figure/themes/change.pdf} 
        \caption{Sub-themes of ``Triggers''}
    \end{subfigure} %
    \begin{subfigure}[t]{0.49\textwidth}
        \centering
        \includegraphics[width=\columnwidth]{figure/themes/quality.pdf} 
        \caption{Sub-themes of ``Quality Assurance''}
    \end{subfigure} %
    \begin{subfigure}[t]{0.49\textwidth}
        \centering
        \includegraphics[width=\columnwidth]{figure/themes/issues.pdf} 
        \caption{Sub-themes of ``Challenges''}
    \end{subfigure} %
    \begin{subfigure}[t]{0.49\textwidth}
        \centering
        \includegraphics[width=\columnwidth]{figure/themes/wishlist.pdf} 
        \caption{Sub-themes of ``Desired Tool Support''}
    \end{subfigure} %
    \begin{subfigure}[t]{0.49\textwidth}
        \centering
        \includegraphics[width=\columnwidth]{figure/themes/llm.pdf} 
        \caption{Sub-themes ``Attitudes Towards LLMs''}
    \end{subfigure} %
\caption{Number and percentage of codes associated with each theme and sub-theme inferred from interview transcripts.}
\label{fig:interview_quant} 
\end{figure}
\end{comment}

\subsubsection{Triggers}\label{sec:rq1_rq2_interview_results_change} 

The primary test maintenance triggers discussed in interviews are summarized as sub-themes in Table~\ref{tab:interviews_change}.

\begin{table}[!t]
    \centering
    \scriptsize
    \caption{Overview of sub-themes of ``Triggers'', with description and number of codes (and percentage of total for that theme) that correspond to that sub-theme.}
    \begin{tabular}{|l|p{9cm}|l|}
    \hline
       \textbf{Sub-Theme}  &  \textbf{Description} & \textbf{Num (\%)}  \\ \hline 
        Production & Changes on the source side cause changes in tests. & 88 (68\%)\\ \hline
        Refactoring & Should not change functionality, but can still induce test changes. & 15 (12\%) \\ \hline
        Bug & The discovery of a bug leads to changes in tests. & 12 (9\%) \\ \hline
        Tech Stack & Changes in the tech stack necessitate changes to tests. & 8 (6\%)\\ \hline
        Testability & Test and source code are changed to be easier to test. & 6 (5\%) \\ \hline
    \end{tabular}
    \label{tab:interviews_change}
\end{table}

\smallskip\noindent\textbf{Production Code:} Naturally, the most prominent reason for test maintenance is a change in the source code. %Naturally, if functionality is changed or new functionality is introduced, the test suite must generally evolve as well. 
Participants, in particular, described test maintenance occurring after new feature implementation, changes in requirements, modifications to code methods, and changes to conditional statements---all of which were also derived in Section~\ref{sec:rq1_review_results}. 

%\interviewquote{%I would say that 
%Production is driving us to change testing. I mean both finding bugs and adding features. Those two really drive up the testing.}{P2}

\smallskip\noindent\textbf{Refactoring:} The source code often evolves as a result of refactoring: 

\interviewquote{Your first try at something is very rarely your best attempt, so generally the same piece of code might be rewritten three or four times, perhaps with different technology or with a different architectural viewpoint, or just in terms of code complexity.}{P1}

Participants noted that it is difficult to completely separate logic and code. %Though ``functionality'' remains unchanged, how it is tested may require modification. 
Refactoring at a small scale---e.g., to improve performance of a function---should ideally not require test maintenance. However, large-scale refactoring, such as a change to the core architecture, may require maintenance, as methods and classes may no longer exist  or there may be interface changes. 

\smallskip\noindent\textbf{Bug:} Naturally, if a bug is discovered, the source code will evolve. This may induce evolution in the test suite as well. For example, developers may discover that an area of the code is under-tested:

\interviewquote{It could be that we noticed something in production not working the way it should. So we locate, OK, where is this code? And then we might connect to, oh, isn't this tested? This logic should work like this, but it works like this. And then we notice that, OK, we don't cover this in tests.}{P2}

\noindent Participants also discussed cases where tests were added to the suite to reproduce a bug, as well as cases where bugs were discovered in the test themselves. 

\smallskip\noindent\textbf{Tech Stack:} Changes to the tech stack---e.g., programming languages, libraries, testing frameworks, or build systems---can also trigger test maintenance. An unstable tech stack can intensify test maintenance, and may constrain the forms and quantity of testing performed. 

%\interviewquote{And then you have a limited tech stack that you can actually use.
%... and these things also change over time. Components that you might have been able to use yesterday, you might not be able to use anymore tomorrow because the licensing requirements change, for example. So a lot of that also impacts the changes that we need to make in our code base in order to keep things running.}{P1}

\smallskip\noindent\textbf{Testability:} At times, the source code is modified to improve its testability. The test suite must be adapted to these changes---e.g., new forms of testing may now be possible. For example, one participant discussed making components more modular or mockable:

\interviewquote{... there are  cases where code is not really testable. Let's say we have a structure where we can't mock stuff ... And then I would say sometimes testing is driving changes in production. So we try to make [a component] more modular.}{P2}

%This sub-theme in particular demonstrates that changes in the test suite should go beyond making sure that tests pass and have high coverage.
%Instead, the sub-theme shows that the complexity and understandability of the test suite should also be a driving factor in modifying it.
%concerns the quality of the test suite and the test maintenance activities that are being, and will be, performed.
%It therefore is quite closely connected to the Effective Test Suite sub-theme from the \textit{Quality Assurance} theme (see Section \ref{sec:rq1_rq2_interview_results_ensure}) in how the values can serve as reasoning to make changes in the test suite itself.

\subsubsection{Quality Assurance} 
\label{sec:rq1_rq2_interview_results_ensure}

\begin{table}[!t]
    \scriptsize
    \centering
    \caption{Overview of sub-themes of ``Quality Assurance'', with description and number of codes (and percentage of total for that theme) that correspond to that sub-theme.}
    \begin{tabular}{|l|p{8cm}|l|}
    \hline
       \textbf{Sub-Theme}  &  \textbf{Description} & \textbf{Num (\%)}  \\ \hline 
       High Coverage & Execute all code to ensure thorough testing. & 46 (22\%)\\ \hline
        Preventative Measures & Finding faults and stopping them from occurring. & 33 (16\%) \\ \hline
       Company Culture & Ensure testing activities are valued and prioritized. & 37 (17\%)\\ \hline
       Insight About Change & Ensure understanding of a change before implementing it. & 28 (13\%)\\ \hline
       Test Logic & Tests should not only test the code, but the logic it represents. & 21 (10\%)\\ \hline
       Effective Test Suite & Tests should be efficient, up-to-date, and relevant, to keep the test suite from growing needlessly. & 20 (9\%)\\ \hline
       Regression Testing & Ensure changes do not break unchanged code. & 17 (8\%)\\ \hline
       Tool Usage & Tools help in discovering faults and increasing code quality. & 10 (5\%) \\ \hline
    \end{tabular}
    \label{tab:interviews_quality}
\end{table}

This theme encapsulates how interviewees ensured test quality, during test creation and maintenance. The sub-themes are explained in Table~\ref{tab:interviews_quality}. %Several of these sub-themes relate to how the testing process is conducted to ensure faults are detected before deployment. The sub-themes also describe the importance of having a high quality test suite to ensure that code and test maintenance and evolution are as painless as possible. 

\smallskip\noindent\textbf{Preventative Measures:} Participants strongly stressed the importance of ensuring code and test quality as a means of ensuring faults are avoided or removed. Participants state that it is better to have a rigorous and time-consuming testing process than to fix faults after the code has been deployed to customers.
%\interviewquote{It's less work to prevent an error than it is to go and fix the error.}{P1}
One method of ensuring quality is test-driven development, where tests are written before the source code:

\interviewquote{I start with the test and then build the code based on that and, therefore, it does definitely change the approach that I take to testing in general... 
%Because the test code isn't written after the feature is written. 
%The test code is part of the feature, 
every time I put down two lines of code, two lines of test code goes with it. You build it piece by piece by piece by piece by testing and testing and testing continuously.}{P1}
\noindent Other preventative measures pointed out by participants include targeting high test coverage, regression testing, code reviews, and collaborating with other teams that use the same code. 

Participants also stressed the importance of careful test maintenance decisions. One discussed a case where a test was removed from the suite, which led to a fault being missed: 

\interviewquote{After that, we have become more cautious and careful when we make the decision to remove test cases ... %because there are ways to prevent similar faults 
%because 
we are only part of the CI flow and there are other CI flows.
Also, each area is testing different scenarios and when we are thinking of removing something, we should always check with the other parts of the CI flow if they have coverage. Then it's probably a lower risk for us to remove it.
}{P7}

\smallskip\noindent\textbf{High Coverage:} Two of the major preventative measure taken by the participants were to ensure %that testing is thorough. Two discussed ways of measuring ``thoroughness'' were 
high structural coverage and high input and output diversity in the tests. 
%\interviewquote{I want to know that every single path through the code is covered. Otherwise, I get nervous.}{P1}
A participant noted that, if coverage measurements are used, it becomes more obvious when the test suite has not co-evolved with the source code: 

\interviewquote{Every line of code should be tested. So when we have just changed a small part of the code then we need to update the test code because otherwise, it complains that not all the lines are covered in test.}{P3}

\smallskip\noindent\textbf{Company Culture:} It is important that testing is valued. % within the company culture. 
%\textit{It's very rare that we that we have to actually address bugs.}
A team that takes pride in code quality will need to edit that code less often. Such teams will also be open to change and feedback.
%One interviewee noted: 
%\interviewquote{I think we are pretty good at testing now. We have had problems with that before ... but I think we're at the point that we actually prioritise it.}{P2}

\smallskip\noindent\textbf{Insight About Change:} Participants discussed the importance of understanding why a change must occur within source or test code before implementing it---whether to fix a bug or perform a planned change.  Gaining an understanding reduces the time to perform maintenance or evolution. %As one interviewee put it when describing how to update a test case: 
%\interviewquote{So first you have to identify: what am I changing and why? Then maybe doing it is not that hard.}{P2}

%Taking the time to understand the fault and 
Explicitly adding a report to the issue tracking system allows the team to match the issue with the right developer, as each have different competencies:

\interviewquote{The next step is always to have some kind of ticket.
Because you, the person that picks it up, is not necessarily the best-suited person to actually complete the task. Make the fix so everything goes through a ticket in the backlog ... And then the next step would be for somebody in the team to pick it up. They will then do a little bit more of a deep dive.
Try and look at the logs. Try and see why the behaviour is what it is.}{P1}

\smallskip\noindent\textbf{Test Logic:} It is important to design tests based not on the code, but on the logic that is the code is supposed to  represent. Participants emphasized that code coverage is not enough, but that tests should also be designed to show that all possible outcomes of a function are demonstrated and that the code fulfills the requirements. They also stressed the importance of isolating individual logical elements when testing: 
\interviewquote{So, we say that first step is to be able to run the method, and this might mean that we need to mock away some API calls, things like that, because we do not want it to actually affect anything outside.}{P2}

\smallskip\noindent\textbf{Effective Test Suite:} Participants discussed their views on an ideal test suite, and how that view influences test maintenance. Each test should be meaningful, and tests should regularly be evaluated to ensure they are up-to-date, relevant, and of high quality.

\interviewquote{In our team, we evaluate the statistics of our tests. How many product faults are they catching? And in that evaluation, we try to find out [which test cases] are not really performing well and are not very efficient. And then we try to remove them.}{P6}

It can be easy for a suite to become bloated over time. Therefore, a significant element of test maintenance is ensuring that irrelevant or redundant test cases are removed. %Ensuring that the test suite is lean and relevant reduces the time spent waiting for feedback and clarifies the debugging process. 

\smallskip\noindent\textbf{Regression Testing:} Code is naturally interconnected, and participants placed importance on ensuring that integration with new code does not affect code that has passed testing. %Code is naturally interconnected, and it is critical to understand how a change in one part of the source code affects another to prevent unintended side effects. 
Regression testing is one means of ensuring that unintended changes are detected quickly. 

%\interviewquote{We then run automated tests, make sure that the fixes haven't broken anything else in the delicate ecosystem.}{P1}

\smallskip\noindent\textbf{Tool Usage:} Two different types of tools were described that help ensure quality---static analysis tools (e.g., linters) and automated testing. Code quality and coverage checks are performed as part of continuous integration. 

%\interviewquote{If I'm breaking anything in the legacy code, I will notice it because the CI tells me something is wrong.}{P8}

\subsubsection{Challenges} 
\label{sec:rq1_rq2_interview_results_issues}

This theme collects test maintenance challenges affecting test maintenance, summarized in  Table~\ref{tab:interviews_issues}.

\begin{table}[!t]
    \scriptsize
    \centering
    \caption{Overview of sub-themes of ``Challenges'', with description and number of codes (and percentage of total for that theme) that correspond to that sub-theme.}
    \begin{tabular}{|l|p{7.5cm}|l|}
    \hline
       \textbf{Sub-Theme}  &  \textbf{Description} & \textbf{Num (\%)}  \\ \hline 
       Growing Test Suite & As test suite grows, harder to efficiently interact with. & 16 (21\%) \\ \hline
       Hardest Task & Opinions on hardest tasks in test maintenance. & 15 (20\%) \\ \hline
       Proliferating Problems & Interlinked problems, such as dependencies between test suites, require more effort to fix. & 12 (16\%)\\ \hline
       Time & Test maintenance is de-prioritized when time is short. & 9 (12\%) \\ \hline
       Missing Coverage & Coverage may be reduced during maintenance, or lack of coverage may trigger maintenance. & 9 (12\%)\\ \hline
        LLM Not Good Enough & Help from LLMs feared to be not good enough to justify the effort required to implement suggestions. & 6 (8\%)\\ \hline
       Unfamiliar with Code & It is harder to find and change unfamiliar test cases. & 5 (6\%)\\ \hline
       Technology Limitations & Tech stack limits possible solutions. & 4 (5\%)\\ \hline
    \end{tabular}
    \label{tab:interviews_issues}
\end{table}

\smallskip\noindent\textbf{Hardest Task:} Participants were evenly split on which test maintenance task was hardest, between identifying how the suite must evolve, modifying tests, adding new tests, and implementing the source change itself. Regarding identifying the change being the hardest task:
\interviewquote{Something is wrong, either with the functionality or the test. 
So first you have to identify what am I changing and why, and then maybe doing it is not that hard. Because then you have these first steps, right? 
... the manual work there is already done.}{P2} 

\noindent Participant P1 thought writing new test cases was harder: 
\interviewquote{Updating  test cases... 
The big thing is the logic is generally there and it's usually just one exception that you found, so it's one specific niche thing.
If you're writing new test cases, you really need to think about all of the ways  things can go wrong, and you need to look meticulously at every path through the code and make sure that you've covered it and have two or three different outcomes that could come out. If you’re updating the test code, that's generally quite small. 
You know what needs to change.}{P1}

\smallskip\noindent\textbf{Growing Test Suite:} Large test suites are harder to manage and understand, and have a long execution time. Test suites grow naturally over time, and continue growing during test maintenance, if care is not taken to remove outdated tests or to modify tests instead of simply adding new tests.

%\interviewquote{By adding test cases it will take a longer time to execute, which is a drawback. Even though we test more, it takes a long time and that is one problem that it takes time when we deliver code ... to go through all the test cases}{P3}

\interviewquote{We don't want to always put in new test cases, because then that would mean we exponentially grow. What's happening is that the maintenance of the test code is as huge as actually developing the product.}{P8}

\smallskip\noindent\textbf{Proliferating Problems:} Multiple elements of both source and test code can be interdependent, which can affect test maintenance. %Naturally, a change to the source code affects the test cases that call that code (or that call dependencies of the changed code). Less obvious is that 
Individual tests are also often linked through their collective contribution to the overall test suite. Changing a test may increase the probability that the suite misses certain types of faults, or may reduce coverage of certain outcomes.

\interviewquote{If you change the tests you could actually be introducing or reintroducing bugs. 
And you could very easily be changing functionality that you didn't intend.}{P1}

The effect of a change on the test suite should be understood in advance. This need introduces hesitance to perform updates and increases the time and effort of test maintenance.  

\smallskip\noindent\textbf{Time:} Participants noted that, when a deadline is approaching, test maintenance activities may be delayed or de-prioritized. A second time-related challenge raised by participants is that testing and maintenance require more manual effort than would be preferred, reducing the total amount of testing or  maintenance that can take place.

\interviewquote{We have quite a robust, arduous process in place when making a commit, a change ... The impact that I think this has is it's a very manual process for us and it slows down the time that it takes and it is a lot of effort.}{P1}

\smallskip\noindent\textbf{Missing Coverage:} Changing or removing a test may result in a loss of coverage. When maintenance is performed, it is important to ensure that the overall level of coverage remains the same or is improved. 

%\interviewquote{A consequence could be that when we remove one test, make changes in the test suite, and then add something else, then the coverage is gone.}{P6}

\smallskip\noindent\textbf{Unfamiliar with Code:} As the project evolves, it becomes more difficult to understand the source code, increasing the difficulty of test maintenance.
%\interviewquote{I wouldn't say dark parts of our code, but like you know, places where we don't really work. So, I don't really know how they work.}{P2}
Instead of changing tests, new tests are added, unnecessarily enlarging the suite. Large suites tend to have high test coverage. However, this can mask problems within the suite. Larger suites are also more difficult to understand. 

\smallskip\noindent\textbf{Technology Limitations:} The tech stack can limit the forms of testing and  maintenance that can take place. For example, there may be requirements on the specific version of a programming language, testing framework, or code library used. Developers may also not be able to use certain tools or frameworks due to privacy or security concerns. 

%\interviewquote{We're limited in terms of what tech we can use, there are certain products that we can't use for technical or for security reasons and so on.}{P1}

\smallskip\noindent\textbf{LLMs Not Good Enough:} Some participants expressed concern that LLMs would not be able to offer sufficiently good or trustworthy performance during testing or test maintenance, and that it would require significant effort to correct their mistakes. They believed that testing could not be fully automated, and would still require significant human oversight. 

\interviewquote{You still have to tell it ´This isn't exactly what I meant. I need this' so you still need that dialogue back and forth.}{P1}

%It should be noted that none of the interviewees had tried to use LLMs for test maintenance and that this experience came from other interactions with LLMs. 
Given the importance of testing, skepticism towards automated approaches is reasonable and should be taken seriously. We further discuss participants' attitudes towards LLMs in Section~\ref{sec:rq1_rq2_interview_results_genai}.

\subsubsection{Desired Tool Support} 
\label{sec:rq1_rq2_interview_results_support}

This theme focuses on requests and opinions regarding automated support for test maintenance, using LLMs or other means, summarized in Table~\ref{tab:interviews_wishlist}.

\begin{table}[!t]
    \centering
    \scriptsize
    \caption{Overview of sub-themes of ``Desired Tool Support'', with description and number of codes (and percentage of total for that theme) that correspond to that sub-theme.}
    \begin{tabular}{|l|p{7.5cm}|l|}
    \hline
       \textbf{Sub-Theme}  &  \textbf{Description} & \textbf{Num (\%)}  \\ \hline 
        Generating Code & Generating test code and improving already-written code. & 20 (29\%)\\ \hline
       Tracking/Understanding & Discovering how/where problems might occur in code. & 13 (19\%)\\ \hline
       Control Test Suite Size & Ensure test suite is  relevant/efficient. & 13 (19\%)\\ \hline
       Ensure Coverage & Prevent and find faults by ensuring high test coverage. & 8 (12\%)\\ \hline
       Ensure Quality & Alerting developers to inadequate testing or code quality. & 5 (7\%)\\ \hline
       Automated Execution & Automating execution of tests. & 5 (7\%)\\ \hline 
       Tool/Team Synergy & Tool should fit into the team's workflow. & 3 (4\%)\\ \hline
       Help and Learning & Making it easier to find and understand new information. & 2 (3\%)\\ \hline
    \end{tabular}
    \label{tab:interviews_wishlist}
\end{table}

\smallskip\noindent\textbf{Generating Code:} As writing code is one of the main activities in testing and development, it is unsurprising that the most common request was for code generation.  This help could come in many forms, from generating tests, to augmenting existing tests, to simply producing an outline or template.

%\interviewquote{A tool that can generate boilerplate tests. 
%That would be the ideal for me.}{P1}

\interviewquote{So let's say I have a method and it takes a list and a string. 
That's simple. Maybe it should create the empty list on this method and an empty string or something like that.  Just have something here so I don't have to think about it. %, tab between all the windows.
}{P2}

\interviewquote{It could detect production code changes and then just [generate] all tests that need to be applied.
%Just press a button and it will generate all the test cases.
That would be great.}{P3}

%\interviewquote{[...] you can't because the files are big and there are a lot of different files. 
%You cannot hand-write it and you can't generate it manually. 
%You have to have some automation tools [...]}{P7}

\smallskip\noindent\textbf{Tracking/Understanding Problems:} There is a desire for tools that detect potential issues in source or test code---a fault, bad practice, or another anomaly---explain the issues, and track whether they have been corrected. 

If a failure occurs, the tester may want to know the source to help in debugging and pritorization:

\interviewquote{We have a tool that will say [...] it a product failure or an environment failure %because that’s something happening here in the lab. 
... We don’t want to spend time on the environment problems, just fix them. We want to spend time on product failures; for this purpose, it is important that this tool has a very good success rate.}{P8}

A developer may also want to know if a failure reappears. Tools could monitor for known failures or anomalies:

\interviewquote{If something's gone wrong before then you don't want to repeat it and you want to be alerted if it goes wrong again.
We create those dashboards manually and then create alerts. I can easily see how a generative AI, for example, could go and look at the data set and see, OK, this is normal behavior ... %, this is what I expect 
and then flag what goes wrong.}{P1}

Participants pointed out that such tools have immense value as significant effort is spent on finding the root causes of, and correcting, failures. 

%\interviewquote{What is hindering us from becoming even better is the time it takes for us to actually correct a failure ... How do we correct the failure? Quick problem identification is key to improve this...}{P8}

%One way a tool could help with tracking and understanding problems is to identify locations where tests are insufficient or may need to be updated.

\smallskip\noindent\textbf{Control Size of Test Suite:} Participants sought tool support for detecting tests that could be removed without losing quality, whether redundant or because the tested code has not been modified for a long time. 
%\interviewquote{There is also a limitation in how long testing can be done.
%... %We allow our SDC to run for 60 minutes. 
%So in that case, we also think that because a test case is quite old and not really catching a lot of product fault, then we, ourselves, decide to remove this one. 
%[...]%and accommodate the new SP request
%}{P6}
This could be done by a tool providing information on how tests overlap in code or outcome coverage or by identifying tests that are potentially out of sync with the code. 

%\interviewquote{Maybe we can use this AI or machine learning to automatically select test cases for changes.}{P7}

\smallskip\noindent\textbf{Ensure Coverage:} Since having high coverage is recognized as important, getting help achieving coverage is important as well. %As attaining high coverage is time-consuming, multiple interviewees expressed a desire for tool assistance.
%\interviewquote{Just having something that makes sure that our coverage is better there, then we know that we can handle different cases in it.
%That would increase my confidence in the code itself, I think.}{P2}
A toool could simply generate tests. However, participants also suggested that tools could augment the input in existing tests.

\smallskip\noindent\textbf{Ensure Quality:} Participants desired tools that detect  problematic development or testing practices and make targeted suggestions, reducing the need for test maintenance. For example: %one participant stated that test coverage may be higher from the start if a tool would alert than to deficiencies that should be corrected: 

\interviewquote{To increase coverage, we could have something telling us coverage is bad and maybe forcing it a little bit more on us. You run Pytest and it passes and then you're happy usually, but you don't look at the coverage report.}{P2}

\noindent Another requested a  pre-check for potential integration issues: 

\interviewquote{I want to have something here that can help the developer to actually pre-check the feature interaction.}{P8}

\smallskip\noindent\textbf{Automated Execution:} Although aspects of test execution are already automated, there are still steps that require manual execution, such as running static analyses or locally executing unit tests before committing. Particularly taxing are cases test execution must be performed manually. Participants sought further automation of the testing process, including assistance with transforming manual tests into automated tests. 

%\interviewquote{That would actually be the ideal system that would look at the interfaces of your different methods that you write and then give you a better idea of whether you are actually covering all of the different scenarios in your tests or even run automated tests on that.}{P1}

\smallskip\noindent\textbf{Help and Learning:} Participants expressed a desire to mitigate knowledge or experience gaps with tool support---e.g., help learning how to write tests in a new language or framework or tips on how to make code more testable.

\interviewquote{It would definitely cut down a lot on development time if you have something that can give you starting points, especially when it comes to using new technologies.}{P1}

\smallskip\noindent\textbf{Tool/Team Synergy:} The way of working in a team has a large effect on how effective any activity---not just test maintenance---can be. Incorporating a new tool requires consideration that  current ways of working are not disrupted, and quality is not made worse because of the disruption. 

\interviewquote{We have a structure to creating tests right?
We start by trying to run the method. We check the output and maybe for the first step at least you're covered.
The assertions, maybe, I want to do myself because I'm not expecting the AI to understand my method...
I don't think it can understand that, but maybe understanding the basics of how our pipeline works. %We create the method, we have a test on it, we always instantiate the method.
}{P2}

\subsubsection{Attitudes Towards LLMs} 
\label{sec:rq1_rq2_interview_results_genai}

Participants' opinions on the use of LLMs in development could broadly be categorized (Table~\ref{tab:interviews_llm}) as skepticism, unfamiliarity, or positivity. Many participants held a mixture of opinions, e.g., both positive towards trying LLMs while skeptical about certain aspects of their use. These attitudes are an important way to gauge the feasibility and desirability of implementing LLM-based tool support for test maintenance. 

\begin{table}[!t]
    \centering
    \scriptsize
    \caption{Overview of sub-themes of ``Attitudes Towards LLMs'', with description and number of codes (and percentage of total for that theme) that correspond to sub-theme.}
    \begin{tabular}{|l|p{9cm}|l|}
    \hline
       \textbf{Sub-Theme}  &  \textbf{Description} & \textbf{Num (\%)}  \\ \hline 
    Skeptical & Do not think LLMs can efficiently help with software development. & 14 (50\%)\\ \hline
    Unfamiliar & Have not used LLMs, either due to lack of opportunity or interest. & 9 (32\%)\\ \hline
    Positive & Like the idea of getting help from LLMs. & 5 (18\%)\\ \hline
    \end{tabular}
    \label{tab:interviews_llm}
\end{table}

\smallskip\noindent\textbf{Unfamiliar:} A few participants were unfamiliar with LLMs.
%\interviewquote{We haven't used it in our daily job [...]}{P7}
At the time, this mainly stemmed from restrictions on their use as part of their work at Ericsson due to security, privacy, and copyright concerns. Although Ericsson has now approved some LLMs, at the time, many employees had not used them at work. Some participants also expressed that a lack of interest.

\smallskip\noindent\textbf{Skeptical:} Some expressed skepticism about the abilities of LLMs. This could stem from a bad experience, things they have heard, a fear of over-hype, or concern that LLMs are being used prematurely to automate certain aspects of development. 

\interviewquote{At the same time, though, am I ready to trust it? 
To cover all of those bases? 
No, I'd still like to have human eyes over it. 
Somebody with some background and the code itself to look at things.}{P1}

%As noted previously, Ericsson must approve LLMs for internal use, and is still in an early experimental phase of LLM adoption. Therefore, some teams have recommended waiting and not using the currently-approved LLMs. 

%\interviewquote{[...] in Scrum master meeting they recommended not to use that [LLM] because it can't track a lot of information [...]}{P5}

\smallskip\noindent\textbf{Positive:} Despite some skepticism, five out of eight interviewees did state that these tools could help out in their work, either currently or in the future.

%\interviewquote{I'm quite keen on the idea of generative AI taking over those types of tasks.}{P1}

\begin{figure}[!t]
    \begin{subfigure}[t]{0.49\textwidth}
        \centering
        \includegraphics[width=\columnwidth]{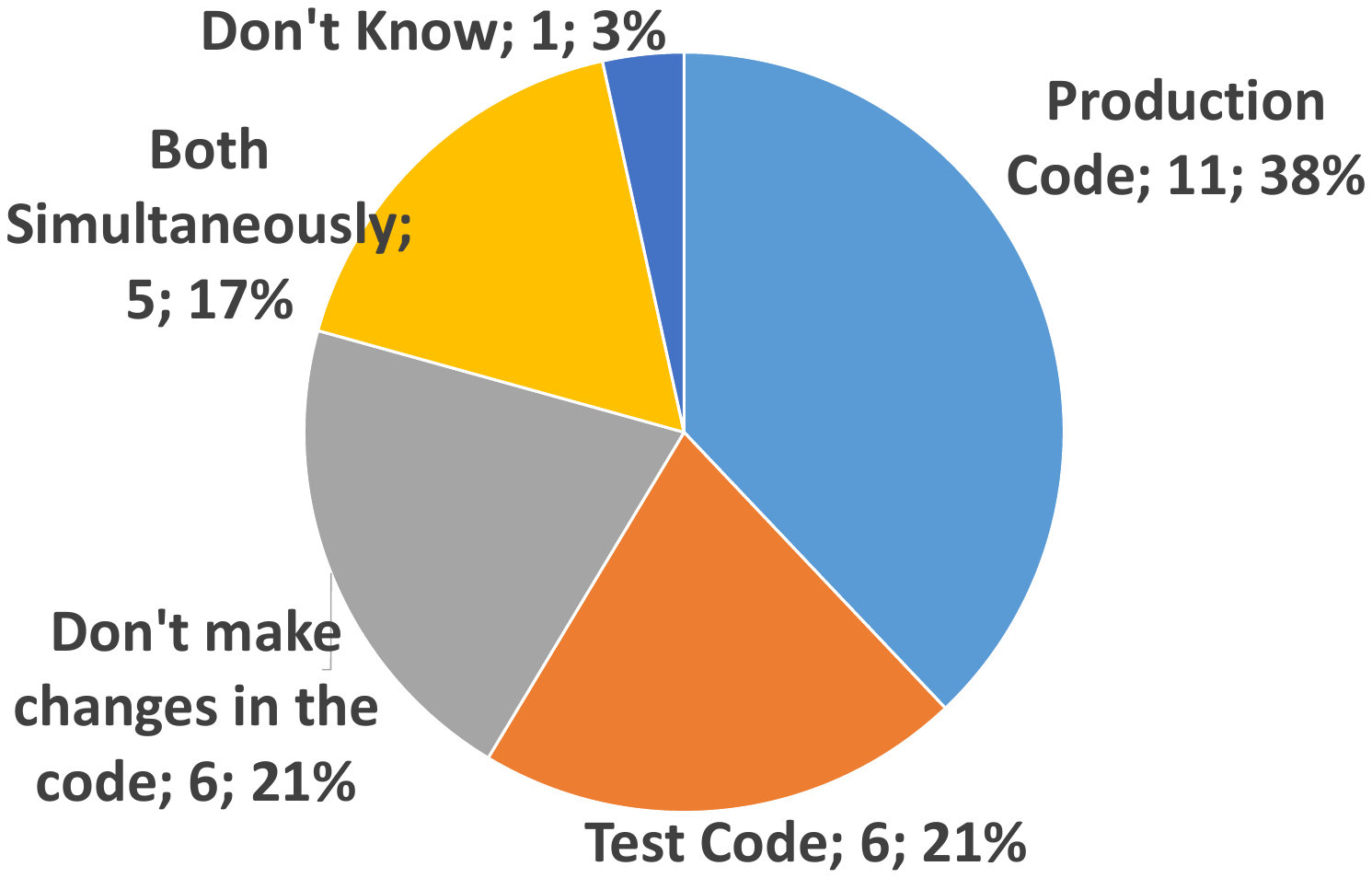} 
        \caption{Do you update source code or test code first?}
    \end{subfigure} %
    \begin{subfigure}[t]{0.49\textwidth}
        \centering
        \includegraphics[width=\columnwidth]{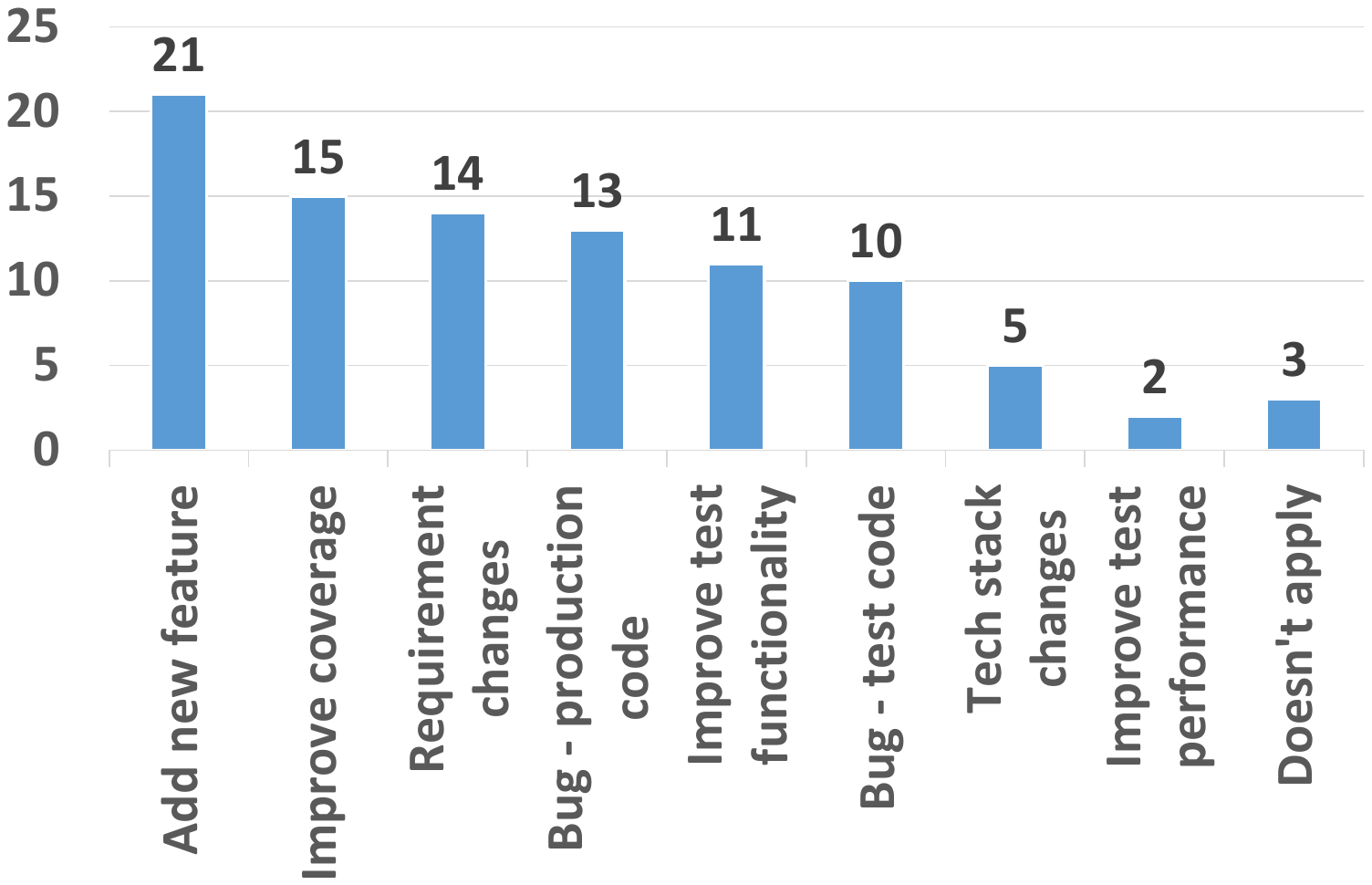} 
        \caption{What reasons have led you to make changes in the test suite or in an individual test?}
    \end{subfigure} %
    \begin{subfigure}[t]{0.49\textwidth}
        \centering
        \includegraphics[width=\columnwidth]{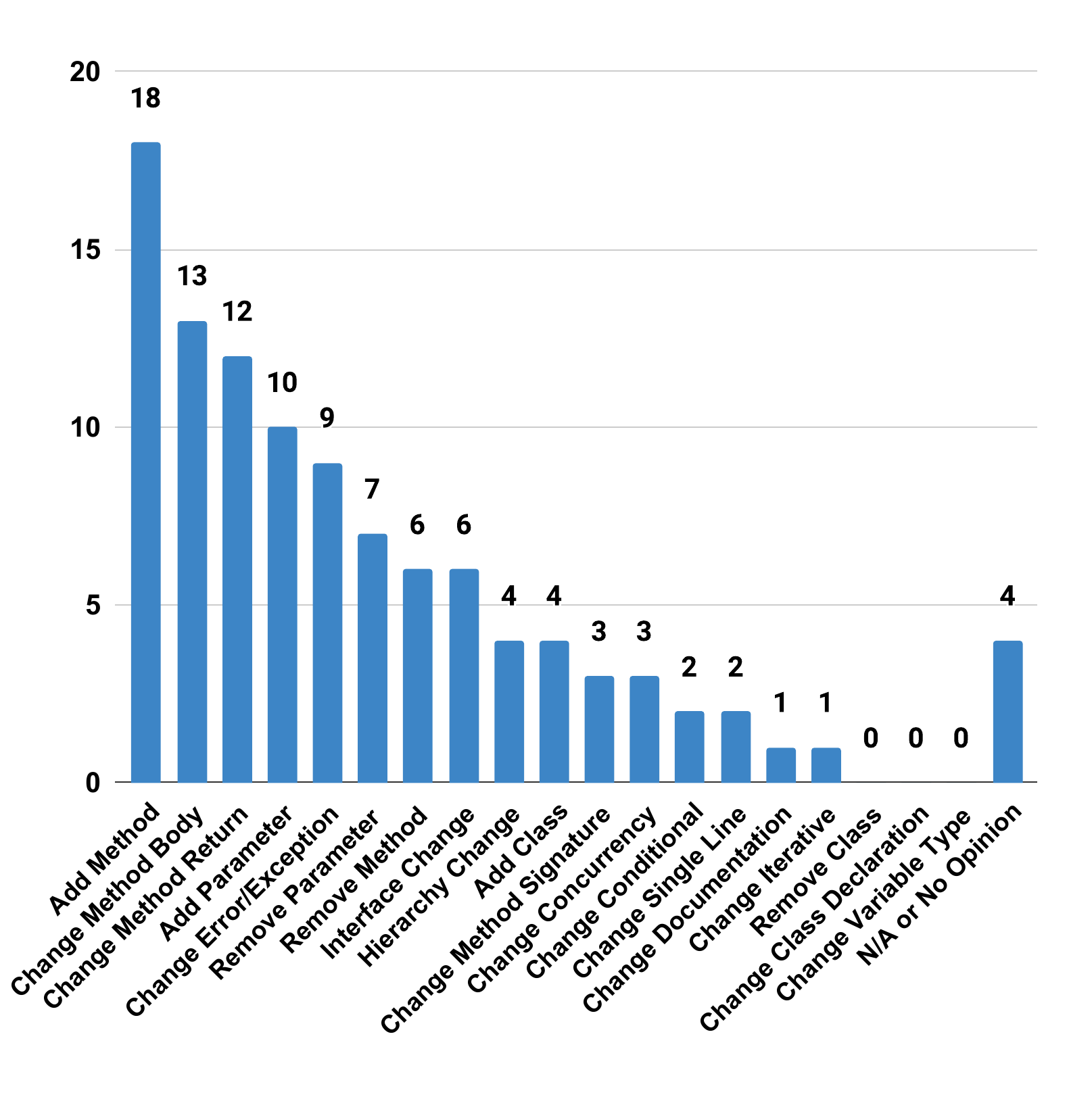} 
        \caption{Which source code changes most commonly necessitate adjustments in associated test code?}
    \end{subfigure} %
        \begin{subfigure}[t]{0.49\textwidth}
        \centering
        \includegraphics[width=\columnwidth]{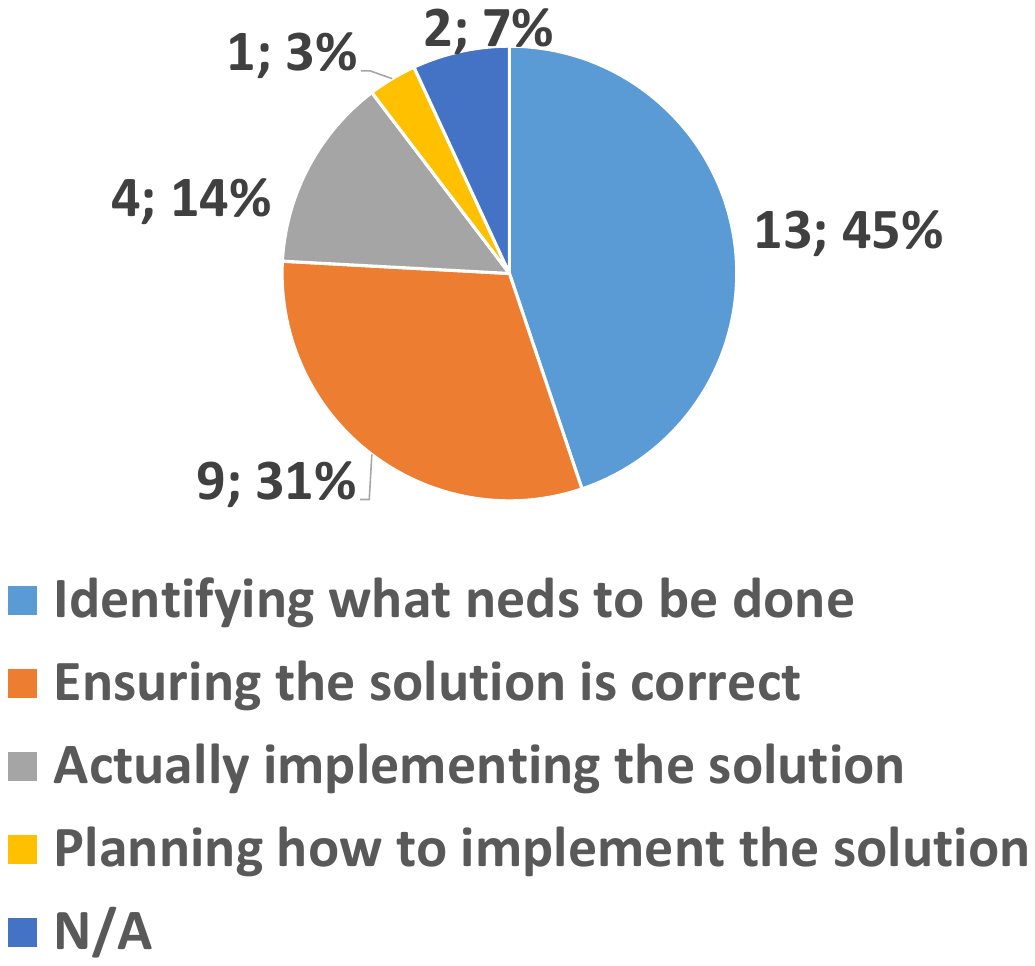} 
        \caption{During the test maintenance process, which step is most effort-intensive?}
    \end{subfigure} %
\caption{Survey responses to questions about the test maintenance process.}
\label{fig:survey_test_maint}
\end{figure}

\begin{figure}[!t]
    \begin{subfigure}[t]{0.49\textwidth}
        \centering
        \includegraphics[width=\columnwidth]{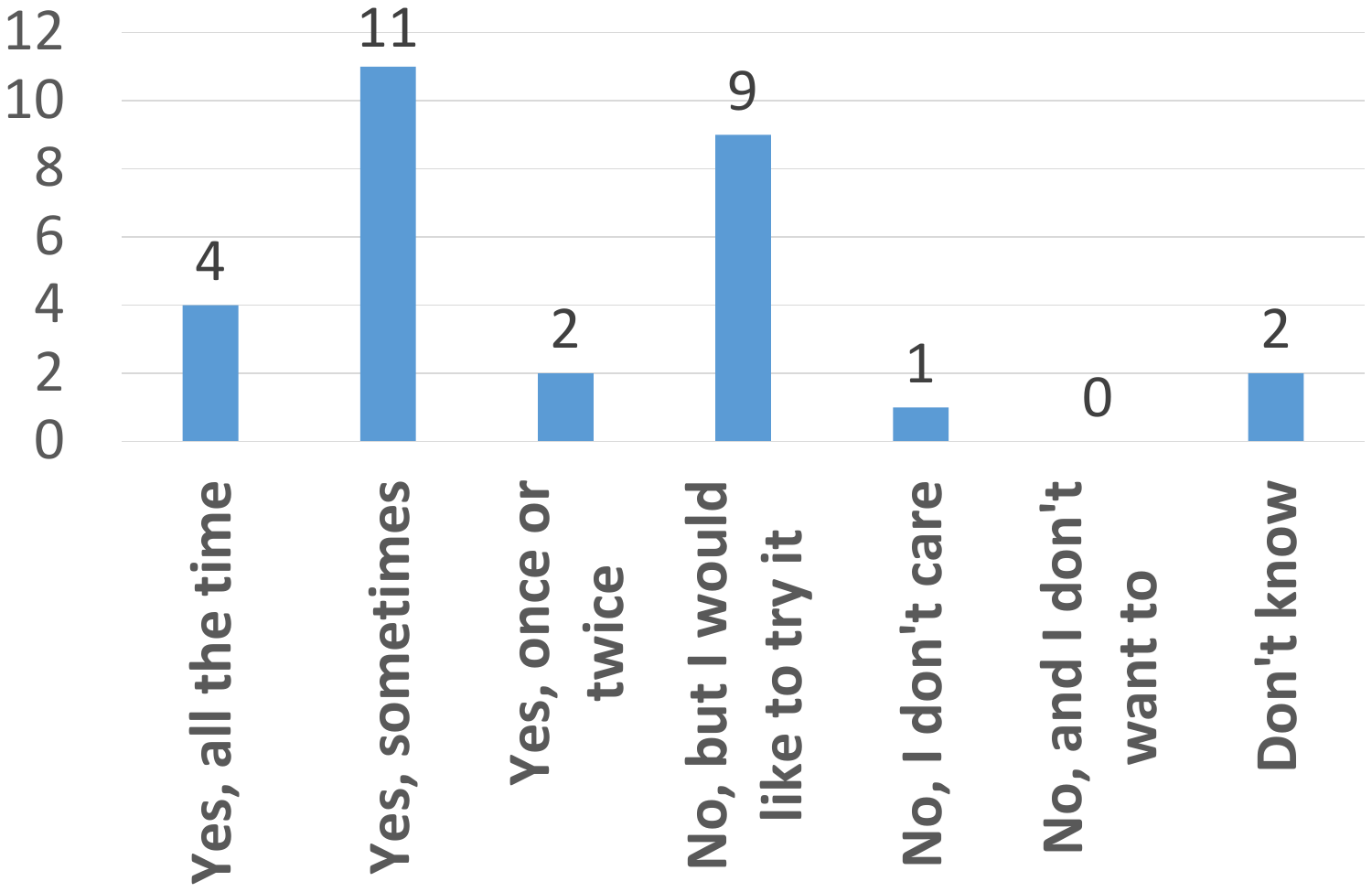} 
        \caption{Have you used generative AI or LLMs?}
    \end{subfigure} %
    \begin{subfigure}[t]{0.49\textwidth}
        \centering
        \includegraphics[width=\columnwidth]{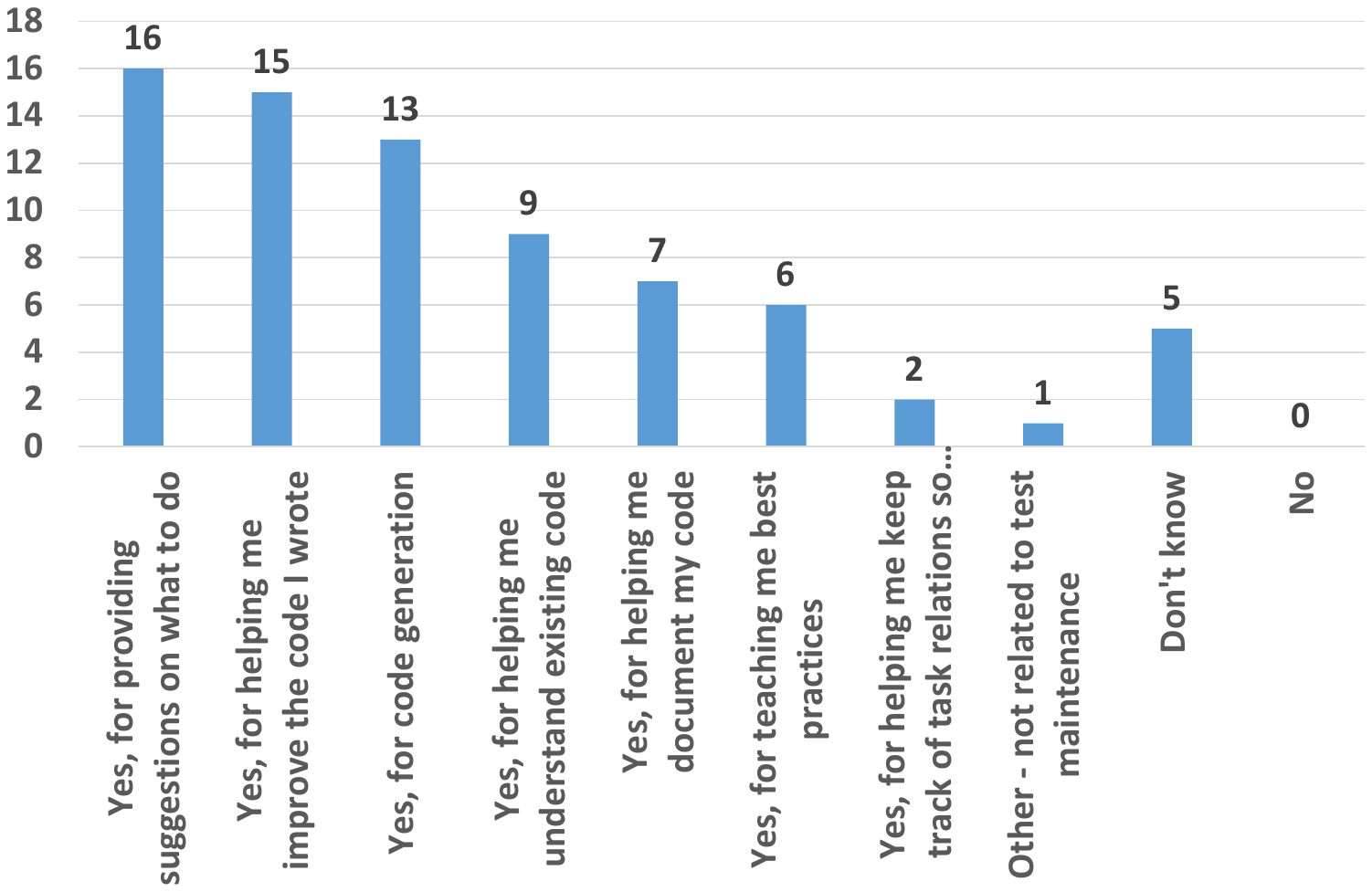} 
        \caption{Would you like the help of generative AI/LLMs for test maintenance activities?}
    \end{subfigure} %
\caption{Survey responses to questions about automation and LLMs.}
\label{fig:survey_llm}
\end{figure}

\subsection{Analysis of Survey Responses (RQ1--2)}\label{sec:rq1_rq2_survey_results}
  
Survey responses related to test maintenance are shown in Figure~\ref{fig:survey_test_maint}. Figure~\ref{fig:survey_test_maint}(a) shows that, while changes to the source code are often made first, there are many cases where tests are changed first or simultaneously. This suggests that test maintenance is not always caused by a code change---i.e., the earlier  list of triggers is not exhaustive. This is further evidenced in Figure~\ref{fig:survey_test_maint}(b), where several common reasons for test maintenance---improving coverage, improving test functionality, fixing a bug in test code---do not require a source code change. %Still, the source code is often updated first, and the most common reason for maintenance is the addition of a new feature. 
In test-driven development, there could also be a change to the tests in anticipation of the change to the source code.

When test maintenance is triggered by a source code change (Figure~\ref{fig:survey_test_maint}(c)), it is most often associated with a method addition, followed by a change to a method body or method return type/value. Three triggers identified in  Section~\ref{sec:rq1_review_results} were not reported by respondents as being among their most common reasons for test maintenance---removing a class, changing a class declaration, or changing a variable or object's type. While these should not be ignored as potential triggers, not all triggers occur with the same frequency, and some source code change may not always lead to test maintenance. 
The most effort-intensive task (Figure~\ref{fig:survey_test_maint}(d)) was identifying what change needed to be made, followed by ensuring that the change was implemented correctly. %Tool support---e.g., the use of LLMs to identify required changes and to identify shortcomings in solutions---could potentially help with both tasks. 

Responses related to automation and LLM use are shown in Figure~\ref{fig:survey_llm}. Figure~\ref{fig:survey_llm}(a) shows that a majority of respondents had tried to use an LLM. Approximately a third still have not tried LLMs, but as shown in Figure~\ref{fig:survey_llm}(b), all respondent expressed a desire to use LLMs as part of test maintenance. The most common requests for assistance included providing suggestions on how to perform test maintenance, improving code that they had written, and generating test or source code. Code generation was only the third most-common desire, suggesting some skepticism and desire to retain control over test suite evolution. 

\subsection{Literature Review on LLM Capabilities for Test Maintenance (RQ2)}\label{sec:rq2_lit_results}

We performed a literature review on the applications of LLMs in software testing and development to, first, suggest test maintenance actions that LLMs may be able to perform and, second, the considerations that must be made when deploying LLMs in an industrial environment. %As we did not find any research specifically on test maintenance, we examined software testing and development more broadly and considered publications that discussed applications with relevance to test maintenance. 

\begin{figure}[!t]
    \centering
    \includegraphics[width=0.8\textwidth]{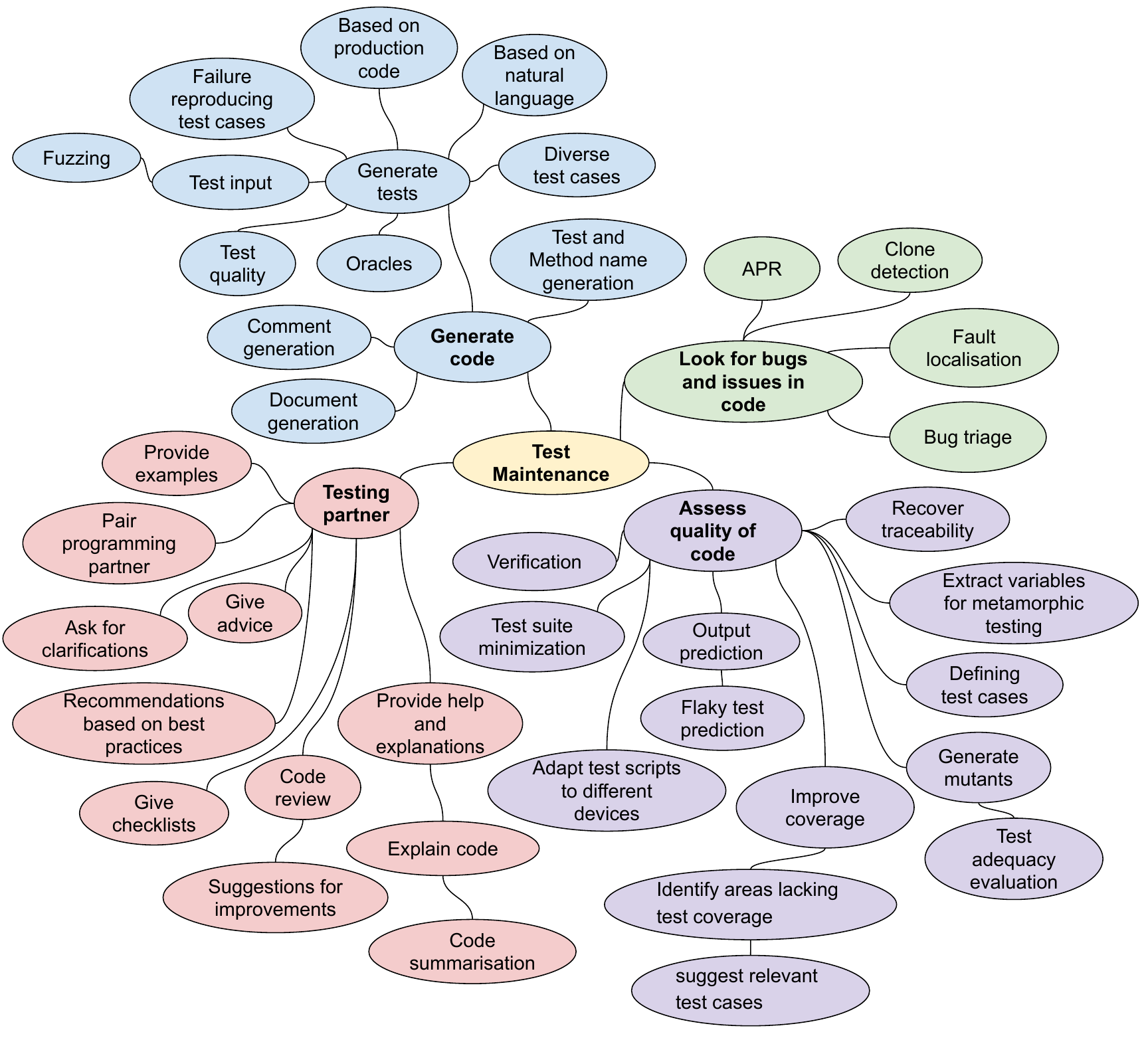}
    \caption{Overview of actions an LLM could take that could assist with test maintenance. Colors indicate clusters of related actions.}
    \label{fig:llm_actions}
\end{figure}

\smallskip\noindent\textbf{Test Maintenance Actions:} Figure~\ref{fig:llm_actions} offers an overview of the actions---suggested by past literature---that an LLM could take that could assist with the test maintenance process. We divide these actions into four clusters, indicated by color, including code generation (blue), testing partner (red), assessing code quality (purple), and discovering issues in the code (green). 

One of the most common uses of LLMs is code generation~\cite{hadi2023large,zhang2023unifying}.
%nazir2023comprehensive, sarkar2022like,xia2023automated,  jiang2023self, fan2023large, hou2023large, ren2023misuse, belzner2023large, chen2023teaching
Naturally, then, LLMs have been used to generate tests (e.g.,~\cite{wang2024software,schafer2023empirical,yuan2023no,lemieux2023codamosa,deng2024large}).
%hou2023large,zhang2023unifying,siddiq2023exploring,fan2023large,bayri2023ai,deng2023large}). 
Creation of tests for new code is part of the maintenance process. LLMs have shown particular adeptness at generating complex input, e.g., for forms~\cite{liu2023fill}
%, zimmermann2023automating 
or deep learning libraries~\cite{deng2024large}. Multi-agent frameworks that iteratively integrate compilation errors have shown promise for reducing hallucinations~\cite{yuan2023no}. LLMs can also generate failure-reproducing tests~\cite{kang2023large}.
%LLM-generated test cases can suffer from hallucinations, resulting in non-compiling code. However, an agent setup where the LLM is automatically provided with compilation errors or where test creation is divided into sub-tasks such as planning and understanding can overcome this challenge~\cite{yuan2023no}. 

LLMs have been used to identify traceability links between natural language and code artifacts~\cite{zhu2022enhancing, lin2021traceability}, and could be applied to link source and test code. LLMs can also explain and summarize code~\cite{hou2023large, zhang2023unifying,nam2024using}, which can be applied to understand how source or test code has changed. 
%nazir2023comprehensive, 

Researchers have explored the use of LLMs as pair programming partners~\cite{ross2023programmer}. %bird2022taking
The user can ask follow-up questions for clarification~\cite{feldt2023towards}. 
The conversational nature of LLMs could allow them to act as a testing partner, offering advice, examples, explanations, and checklists.  
%Test maintenance is a process where a developer requires a deep understanding of both the source and test code, as well as knowledge of good testing practices. LLMs could offer conversational assistance.

LLMs can provide suggestions for code improvement~\cite{hou2023large,zhang2023unifying, lu2023llama}.
%belzner2023large,
Such capabilities could be applied to the test suite. In addition, LLMs can identify additional scenarios to explore, areas lacking code coverage, or flaky tests~\cite{hou2023large,fan2023large,sarkar2022like}.
%identify weaknesses in the test suite through mutant generation~\cite{ibrahimzada2023automated,fan2023large,zhang2023unifying,hou2023large}, 
LLMs can also adapt tests to different devices~\cite{yu2023llm}, extract properties for metamorphic testing~\cite{tsigkanos2023large}, improve test readability~\cite{Gay23:Readability}, and minimize test suites~\cite{fan2023large}. LLMs can generate documentation~\cite{hadi2023large,hou2023large,fan2023large},%,nazir2023comprehensive}
code comments~\cite{Gay23:Readability,geng2024large}, as well as method and test names~\cite{hou2023large,Gay23:Readability}. All of these capabilities could be used to improve a test suite.

LLMs can also be used as part of debugging~\cite{wang2024software,hou2023large} and automated program repair~\cite{hou2023large, xia2023automated}. % zhang2023surveyARP, 
LLMs can also identify issues (e.g., faults, code smells) in source code~\cite{hadi2023large,hou2023large,zhang2023unifying, fan2023large}.
%,sarkar2022like,belzner2023large,nazir2023comprehensive,    li2023hitchhiker, chen2023teaching}. 
%These same capabilities may have applications as part of the test maintenance process as well. 

\begin{figure}[!t]
    \centering
    \includegraphics[width=0.7\textwidth]{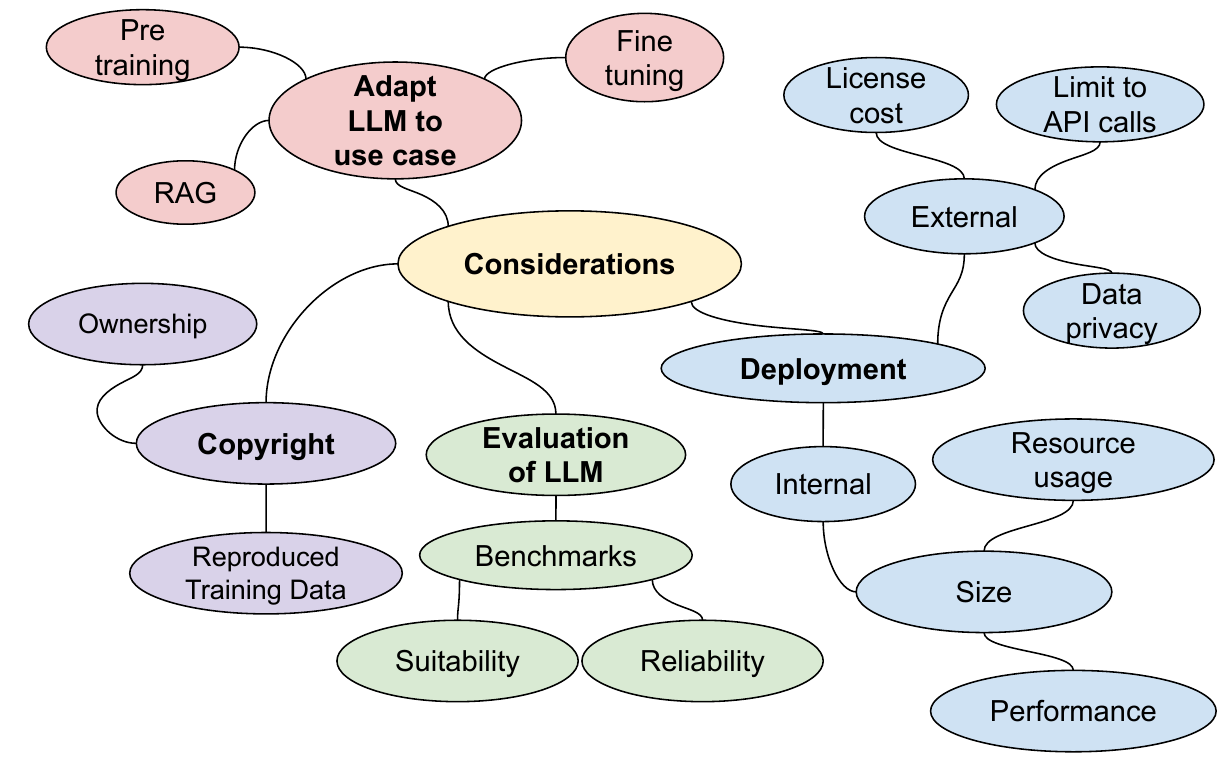}
    \caption{Considerations for LLM deployment in an industrial environment. Colors indicate clusters of related considerations.}
    \label{fig:llm_considerations}
\end{figure}

\smallskip\noindent\textbf{Considerations for LLM Deployment in Industrial Development:} Considerations discussed in past literature are shown in Figure~\ref{fig:llm_considerations}. %There are many different LLMs, with more emerging regularly. These considerations are not exhaustive, but can help guide LLM selection. 

LLMs have varying performance across tasks depending on how they have been trained and tuned, e.g., some models are tuned purely for source code tasks, while others are tuned for a blend of source code and natural language~\cite{roziere2023code}. 
%For example, Roziere et al. noted that---in their study---the Code Llama model performed better than the GPT-4 and GPT-3.5 turbo models for code generation tasks~\cite{roziere2023code}. This was because Code Llama was specifically trained and tuned for source code-based tasks, while the other models were trained on a mixture of code and natural language and were intended for reasonable general performance. Naturally, then, 
The first consideration when selecting a model is the type of task that needs to be performed. 
A model tuned for that task will almost always outperform models of the same size (measured in the number of parameters)~\cite{zheng2023survey}. If many models are appropriate for a given task, evaluations on public benchmarks can be used to perform a comparison---e.g.,  HumanEval~\cite{chen2021codex} for code generation or the LMSYS Chatbot Arena for conversational tasks~\cite{chiang2024chatbot}. However, many software engineering tasks do not have ``standard'' benchmarks yet~\cite{wang2024software}. 

However, benchmark performance may be affected by data leakage~\cite{wang2024software}. %This is especially a risk for widely-used benchmarks or those not designed specifically for LLMs~\cite{wang2024software}. In addition, 
Current benchmarks, such as HumanEval, have also been criticized for having insufficient test suites and for using vague problem descriptions~\cite{liu2024your}. Both could result in misleading performance estimates.
%Benchmark performance is still a valid starting point, but as a result of these factors, stated performance scores may not reflect real-world performance on industrial codebases. 
%Unless a company trains their own model, industrial code should not appear in the training data for the LLM, preventing data leakage. Industrial test suites are also likely to be stronger than ones built for a benchmark. 
In general, the larger the model, the better it performs~\cite{zheng2023survey}. 
Larger models also demonstrate emergent abilities, such as in-context learning~\cite{brown2020language} and step-by-step reasoning~\cite{shanahan2024talking}, that are not always found in smaller models~\cite{zheng2023towards}. However, larger models impose significant computational and energy costs~\cite{wang2024software}. Models can be fine-tuned on an industrial codebase to improve their performance, and smaller fine-tuned models can sometimes outperform larger models~\cite{zheng2023survey}. 

There are also organization factors to consider. If a model is deployed by an external company, the cost of a license and limitations on the number of API calls must be considered~\cite{mandvikar2023factors}. There may also be privacy concerns, related to sending proprietary data to a third party~\cite{wang2024software}. If an organization can afford the computational costs of internal deployment, then that is generally a better option. Internal deployment avoids privacy concerns~\cite{wang2024software}. The organization can also tune models, change parameter settings, and choose when and how to deploy updated models~\cite{wang2024software}.  

An additional concern is copyright. LLM output, code and natural language, is derived from training data. LLMs may output copyrighted text, which could lead to issues if, e.g., that code is incorporated into the organization's  codebase~\cite{li2024diggerdetectingcopyrightcontent}. It is unclear whether an organization can own copyright over code generated by an LLM~\cite{ownership}. %Copyright issues can be mitigated by choosing an open-source LLM where the training data can be inspected. However, open-source LLMs are less common than ``open-weight'' models, where weights are available but not underlying training data~\cite{openweight}. 

\subsection{Proof-of-Concept Evaluation (RQ3)}\label{sec:rq3_eval_results}

\begin{comment}
\begin{table}[!t]
    \small
    \caption{Comparison of results of the prototypes. The highest performance in each measurement is \textbf{bolded}.}
    \label{tab:eval_filtered}
    \centering
    \begin{tabular}{ |l | c | c | c | c | c | c | }
         \hline
          & \multicolumn{3}{c|}{\textbf{Average}} & \multicolumn{3}{c|}{\textbf{Aggregate}} \\ \hline 
         \textbf{Prototype} & \textbf{Recall} &\textbf{ Precision} & \textbf{F1 Score} &  \textbf{Recall} & \textbf{Precision} &  \textbf{F1 Score} \\
         \hline
         Planning (With Summary) & \textbf{0.5859} & 0.1885 & 0.1801 & \textbf{0.3367} & \textbf{0.2596} & \textbf{0.2932}\\ \hline
         Planning (Without Summary)  & 0.5776 & \textbf{0.3257} & \textbf{0.3048} & 0.1867 & 0.2343 &  0.2078 \\ \hline
         Chain (With Summary) & 0.5824 & 0.1310 & 0.1174 & 0.3000 & 0.1863 & 0.2299 \\ \hline
         Chain (Without Summary) & 0.5165 & 0.1211 & 0.0847 & 0.2233 & 0.1138 & 0.1507\\ \hline
    \end{tabular}
\end{table}
\end{comment}

\begin{table}[!t]
    \centering
    \scriptsize
    \caption{Evaluation of prototypes when $> 0$ tests are updated in the ground truth.}
    \label{tab:eval_changed}
    \centering
    \begin{tabular}{ |l | c | c | c | c | c |  }
         \hline
         \textbf{Prototype} & \textbf{Recall} &\textbf{ Precision} & \textbf{Accuracy} & \textbf{F1 Score} & \textbf{F2 Score} \\
         \hline
         Planning (With Summary)& \textbf{0.676} & \textbf{0.275} & \textbf{0.911} & \textbf{0.391} & \textbf{0.523}\\ \hline
         Planning (Without Summary)& 0.412 & 0.213 & 0.908 & 0.281 & 0.347   \\ \hline
    \end{tabular}
\end{table}

\begin{table}[!t]
    \centering
    \scriptsize
    \caption{Evaluation of prototypes when no tests are updated. Only FP and TN are possible in this scenario.}
    \label{tab:eval_unchanged}
    \centering
    \begin{tabular}{ |l | c | c | c |  }
         \hline
         \textbf{Prototype}& \textbf{Avg. FPs/commit} & \textbf{Accuracy} \\
         \hline
         Planning (With Summary)& 5.5 & 0.974   \\ \hline
         Planning (Without Summary)& \textbf{3.7} & \textbf{0.982}\\ \hline
    \end{tabular}
\end{table} 

We evaluated the performance of two prototypes---one that makes predictions based on the raw test code and one that generates and operates on natural language summaries of the test code---over the dataset. This evaluation was performed twice to assess variance. As the results of the two executions were very similar, we present average results. To perform this evaluation, we split the dataset based on two scenarios. The first subset contains cases where one or more tests were updated in the ground truth, while the second contains cases where no tests were updated. 

%It is worth noting the imbalance in the input data between commits with and without test case changes. Among the 54 commits used for evaluation, 32 commits have corresponding ground truth test case changes, containing a total of 535 git diff messages—an average of 16.7 per commit. In contrast, the remaining 22 commits have no associated test case changes and contain only 99 git diff messages, averaging 4.5 per commit. This indicates that input data with ground truth test case changes contains significantly more code changes than data without test case changes.  %We discuss this observation further in Section~\ref{sec:discussion}. 

Table~\ref{tab:eval_changed} presents the average performance for cases where tests were updated in the ground truth. The best performance in each measurement is \textbf{bolded}. The planning agent that makes use of summaries attained the highest performance in all measurements, suggesting a potential benefit from translating the test code to natural language for comparison with the natural language summary of the corresponding source code change. 

Table~\ref{tab:eval_unchanged} presents performance measurements for cases where no tests were updated in the ground truth. For this subset, only FP and TN results are possible, meaning that recall, precision, F1, and F2 score are undefined for this subset. For this subset, the best prototype was the one based directly on the test code. 

The accuracy for both subsets is very high. However, this measurement is skewed by the large number of TN cases. The test suite for the repository is large. In most situations, a very small number of tests will be updated when the source code is updated. Therefore, it is useful to focus on the other measurements to better understand the performance of the prototypes.

In the first subset (Table~\ref{tab:eval_changed}), the recall is relatively high for the prototype based on summaries, but precision is low. In Table~\ref{tab:eval_unchanged}, we also indicate the average number of FP per commit. Together, these results suggest that the primary factor affecting performance is FP---this prototype is more likely to suggest \textit{extra} tests than to miss an important test. 

Because it works with natural language rather than code, the prototype based on summaries makes a larger number of positive predictions (TP and FP) than the prototype based on test code. This improves its performance in situations where tests have been updated in the ground truth, but leads to worse performance in cases where no tests were updated.

\begin{figure}[!t]
    \centering
    \begin{minted}{text}
The following test cases require maintenance due to the code changes:

1. `testParameters()`: Needs to be updated to reflect the new nested structure of the 
   `parameters` field.
2. `testConfigLoads()`: May need to be updated to ensure that the new configuration 
    elements, including the `DEFAULT_KEY`, are being loaded properly."
    \end{minted}
    \caption{An example of a confident prediction that maintenance is required.}
    \label{fig:prediction_confident}
\end{figure}

\begin{figure}[!t]
    \centering
    \begin{minted}{text}
No test maintenance is needed based on the provided code changes. The changes involve 
configuration updates that do not directly impact the existing test cases identified.
    \end{minted}
    \caption{An example of a confident prediction that no maintenance is required.}
    \label{fig:prediction_no_confident}
\end{figure}

\begin{figure}[!t]
    \centering
    \begin{minted}{text}
While no specific test cases were identified as needing updates based on the provided 
information, it is recommended to review and potentially update any test cases that 
interact with the `FlowConfig` class or depend on the behavior of the methods 
`isRenamingEnabledFor`, `getEmptyFileSize`, `getSubdirectoryDatePattern`, 
`getParameterValue`, and `getParameters`. This is due to the introduction of the `project`
parameter and the change in the return type of `getParameters`, which may affect the 
expected behavior of these methods.
    \end{minted}
    \caption{An example of a general suggestion that tests be reviewed.}
    \label{fig:prediction_review}
\end{figure}

\begin{figure}[!t]
    \centering
    \begin{minted}{text}
No existing test cases are directly impacted by this change, but new test cases should be 
created to cover the updated behavior of `getNextTransformation`, specifically focusing 
on scenarios involving different combinations of `eventData.getProject()` and 
`eventData.getFileClassification()`.
    \end{minted}
    \caption{An example of a suggestion that new tests be created.}
    \label{fig:prediction_new}
\end{figure}

The predictions made by the prototype are not strict ``yes'' or ``no'' predictions, but are more nuanced. At times, the prototype will confidently suggest that certain tests should be changed (Figure~\ref{fig:prediction_confident}) or that no maintenance is required (Figure~\ref{fig:prediction_no_confident}). At other times, when less confident, it may suggest that a particular test should be reviewed or make general recommendations about reviewing tests related to particular code elements (Figure~\ref{fig:prediction_review}). Sometimes, it even suggests creating new tests rather than performing maintenance (Figure~\ref{fig:prediction_new}). The languages used in the output also varies, e.g., suggesting that a test ``needs to be updated'',  ``should be updated'', or ``might need to be updated'', based on the prototype's level of confidence.

These general suggestions or even a FP---even if not technically \textit{correct}---may still benefit a tester as they update the test suite. These suggestions may have a role in the brainstorming process, as long as the number of FP remains relatively low. We further investigate the utility of these suggestions in Section~\ref{sec:post_interview_rq3}. 
FN results are more concerning, and in future work, we will explore means of improving recall further.

\subsection{Developer Assessment  (RQ3)}\label{sec:post_interview_rq3}

\noindent\textbf{Usefulness Validation:} Both participants offered generally positive impressions. Although the prototypes did not fully ``solve'' test maintenance challenges, they were described as something that would be nice to have. They noted two main benefits. First, the prototype runs much faster than executing the tests---generally returning a result in under 30 seconds---which means they could be made aware of problems faster. The prototype may also be more reliable for some cases, as a test may still pass the existing test suite despite requiring maintenance. Second, they saw value in a ``second opinion'' beyond the intuition of the human reviewing the test cases. Although a human still made decisions, the prototype offered guidance that could be factored into their decision, increasing their confidence. 

\smallskip\noindent\textbf{Assessment by Commit Author:} In the first session, we asked questions about how the developer performs test maintenance. Similar to suggestions made by the interview and survey participants, this developer primarily relied on two sources of information. The first is their intuition about the codebase, and the second is the results of running the test suite on the updated codebase. However, they noted that these signals are not enough to feel confident that all tests that need maintenance have been correctly identified.

%\interviewquote{
%So you're like you're making, like one change, then you just run the the test suite and then it will maybe produce some errors...}{The developer}

%\interviewquote{
%It's a combination. Some are found by failed test cases in the test suite, some you already have in your mind. If you change something like a sending message, then most probably you need to also check the test case that tests sending message ... However, I am not that confident that all related test cases are found...}{The Developer}

\begin{table}[!t]
\centering
\scriptsize
\begin{tabular}{|l|l|l|l|}
\hline
\textbf{Category} & \textbf{Subcategory} & \textbf{Count} & \textbf{Total} \\
\hline
\multirow{2}{*}{Useful} 
    & Test case needs to be changed (TP) & 7 & \multirow{2}{*}{16} \\
    & Suggestion carries benefits to suite or maintenance & 9 & \\
\hline
\multirow{2}{*}{Not Useful} 
    & Test not relevant to changed code & 10 & \multirow{2}{*}{14} \\
    & Unreasonable explanation & 4 & \\
\hline
\multirow{2}{*}{Other} 
    & Problem with ground truth (TP) & 7 & \multirow{2}{*}{8} \\
    & Unable to understand test case & 1 & \\
\hline
\end{tabular}
\caption{Categorization of FP suggestions by commit author.}
\label{tab:test_case_categories}
\end{table}

In the second session, we presented 38 FPs from three commits. Initially, we presented the name of each test and asked the developer to classify the suggestion as useful or not, and to explain why. We then presented the prototype's full output for each FP and asked them to re-examine their previous classification. The results of this exercise are shown in Table~\ref{tab:test_case_categories}.

Of the 38 FPs, the developer identified 16 suggestions as useful. In seven cases, the developer determined that these cases were actually TP---they had missed these tests during the initial maintenance process. An additional nine suggestions were not correct, but the suggestion was still seen to have some potential benefit, such as suggesting new tests to create.

A further 14 cases were confirmed as FP. In 10 cases, the suggested tests had no relation to the changed code. In the remaining four cases, the explanations from the tool were deemed unreasonable.

Among the remaining eight FPs, seven were actually updated (i.e., were TP), and were missed when creating the dataset. Creating the dataset required determining which test updates were relevant to the code changes. These cases reflect either labeling errors or cases where a test was updated in a later commit. Finally, there was one case where the developer could not understand the test well enough to make a determination.

%found 8 FPs are actually already changed and should be included in the ground truth test cases, this should be due to the annotation mistakes when manually sorting out the ground truth test cases; one test case is found to be badly written and is excluded for evaluation; 10 FPs are actually valid suggestions, which means either the developer missed it when changing test cases, or the output makes sense to let the developer examine on it. The rest 21 FPs were marked by the developer as "invalid", which means he still thinks they are irrelevant to the code change and do not need to be maintained.

%The developer was checking if the reason given by the prototype is correct and substantiated and the suggestion made by the LLM is relevant to the changed code that it will improve (e.g., more coverage) the existing test case.

%After seeing the prototype's output for each FP, the developer re-examined each FP and the corresponding code change, and changed another 6 FPs that are previously labeled as "invalid" to "valid", in total 16 FPs labeled as "valid" now. Among the 16 FPs labeled "valid", 7 FPs should be changed while the rest should be reviewed as the prototype output suggested but do not need to be changed after reviewing by the developer. This results in only 13 FPs left as invalid suggestions out of the original 38 FPs. For the 13 FPs that remained labeled as "invalid", 9 of them the developer thought the test cases does not cover the corresponding code change, 4 of them have unreasonable output. The final result is presented at Table~\ref{tab:test_case_categories}. 

Based on this assessment, we observe that the results in Section~\ref{sec:rq3_eval_results} are potentially conservative. Of the 38 FP cases, 14 were actually correct suggestions---either missed by the developer or during the creation of the dataset. In a further 10 cases, the suggestion was not correct, but still offered some perceived benefit, suggesting tests that should be closely reviewed or newly created. 

%From this we could see that the prototype is already quite helpful, for even the most experienced developer for one specific repository, where it helped the developer to find missed test cases that should have been changed, or other useful suggestions like reviewing test cases that were not checked before or even suggesting creating new test cases to improve the coverage for the existing test suite. We could also sense that a lot of the so-called "FPs" or "ground-truth" test cases are not immaculate, since human mistakes exist all the time, e.g., even the most primary contributor to the repository would make mistakes like missing test cases that should be changed in test maintenance procedures. With the above findings we have, the actual recall and precision number, especially the precision number, is only the lower bound due to human errors, that is another reason why we think recall and F2-score matters more than precision and F1-score in our test maintenance scenario.

The final session revolved around the developer's perceptions of the tool and preferences on how it should be tuned. The developer placed some emphasis on recall over precision:

\interviewquote{
%It is important that more TPs are actually found by the tool.
It's good to have a balance between recall and precision, but if there's a slight decrease in precision with improved recall, this is more acceptable. It's probably better to optimize recall than precision in our use case.
}{The Developer}

The developer positively noted the feedback from the prototype, the suggestions of new tests to create, that explanations were provided even when it indicated that no tests needed to be updated, and that the language employed varied with the prototype's level of confidence. However the developer would like to see more explanation of each suggested test. During the development process of the prototype, we incorporated instructions into the prompt to limit the length of the explanation, based on earlier feedback that indicated that the explanations were too long. This suggests that multiple prompts could be employed based on a users' preference for explanation quantity.

\section{Discussion}\label{sec:discussion}
In this section, we provide answers to the research questions, based on the results of our research activities. We also discuss threats to validity.  

\subsection{Test Maintenance Triggers (RQ1)}\label{sec:discussion_rq1}
    
Our results suggest the existence of both \textit{low-level} and \textit{high-level} triggers that create a need for test maintenance. 

\smallskip\noindent\textbf{Low-level Triggers:} Past research has focused predominately on test maintenance triggers related to individual changes to the source code or documentation. These triggers are summarized in Tables~\ref{tab:triggers_func}--\ref{tab:triggers_line}, and were confirmed during the interviews and survey. 

That low-level changes can trigger test maintenance is intuitive. If functionality evolves, its test must also evolve. If functionality is added or removed, then tests should be added or removed as well. However, the existence of such triggers \textit{does not guarantee} that maintenance is required. All of the commits in the evaluation dataset used to address RQ3 contain source code changes---i.e., low-level triggers---but not all of those changes led to test code changes. In practice, a test will only need maintenance if the test executes the changed lines of code and if the test is somehow no longer in alignment with the the changed source code---e.g., there has been some change in how the code is invoked or a change in its behavior that is detected by the test oracle. Future research should examine the causal relationship between source and test code evolution in more detail to better predict when a trigger will lead to requires maintenance.

%A test suite may not execute changed code, or if it does, it may not detect the change (e.g., due to a weak oracle, low input diversity, or low coverage of alternative paths containing changed code). In such cases, the test suite may not require maintenance---although a lack of need for maintenance, in itself, may be an indication that the test suite is weak. It may also be the case that test maintenance was not performed until a later commit, as we only considered co-evolution that occurred in the same commit. 

%Some of these triggers are also broad enough that they may not be useful for prediction purposes. For example, ``removing functionality'' can describe many different code changes. Because it is so broad, it could be considered a trigger, even for changes that do not necessitate test maintenance.

%Alternatively, it may be that these triggers should be combined with additional data when used for prediction and decision-making. In our prototypes, we did not provide the LLM agents with code coverage information. Instead, the agents made their own determination of traceability between source and test code. Coverage information could ensure, at least, that the changed code is covered by a test. Combinations of triggers and other analyses of source and test code or other artifacts may be needed to determine whether test maintenance is actually required. 

\begin{coloredframe}{blue}
\textbf{Test Maintenance Triggers (RQ1):} We identified 37 low-level changes to functionality, documentation, classes, methods, and single lines of code that can trigger test maintenance in situations where test cases invoke and detect those changes.
\end{coloredframe}

\begin{table}[!t]
\centering
    \scriptsize
    \caption{Description of high-level triggers that may indicate a need for test maintenance.}
    \centering
    \begin{tabular}{|l|p{9cm}|}
        \hline
        \textbf{Trigger} & \textbf{Description} \\ \hline 
        Adding New Feature & Expanding on an existing product by adding new functionality, leading to the creation of new tests and adjustment of existing tests. \\ \hline
        Change in Requirements & Existing feature requirements have been modified, leading to a need to change both source and test code. \\ \hline
        Discovery of a Fault & The discovery of a fault in the code base, either by a developer or documented in an issue report, cause a need to add or modify tests to ensure the fault has been repaired and that it will be caught if it appears again. \\ \hline
        Increase Coverage & The test suite is modified and new tests are added to increase code coverage. \\ \hline
        Refactor Code & The code is refactored or enhanced with the intention to preserve the current functionality while improving quality, e.g., its performance. Both test and source code can be refactored, both can require test maintenance.\\ \hline
        Tech Stack Changes & Changes in technologies used in the development process may require test maintenance. \\ \hline
        Evaluation of Test Suite & Systematic evaluation of a test suite may identify weaknesses and lead to maintenance.\\ \hline
    \end{tabular}
    \label{tab:rq1_high_level_factors}
\end{table}

\smallskip\noindent\textbf{High-Level Triggers:} From interview and survey responses, we identified additional ``high-level'' test maintenance triggers. These triggers represent decisions made as part of the development process rather than individual changes to the source code. For example, there may be a decision to increase the code coverage of the current test suite. There may not have been a change to the source---hence, no low-level trigger---but tests may be added or existing tests may be augmented. Like with the low-level triggers, such decisions do not always guarantee test maintenance, but are frequently associated with it. We present these high-level triggers in Table~\ref{tab:rq1_high_level_factors}.

\begin{coloredframe}{blue}
\textbf{Test Maintenance Triggers (RQ1):} We identified seven high-level development decisions that often trigger test maintenance, including adding features, requirement changes, fault discovery, increasing coverage, refactoring, tech stack changes, and test suite evaluation. 
\end{coloredframe}

\subsection{Applications of LLMs in Test Maintenance (RQ2)}\label{sec:discussion_rq2}

We examined the theoretical applications of LLMs or agents in  test maintenance, considering how the triggers could be acted upon, what actions could be taken based on these triggers and other data signals, and what considerations should be made when deploying LLMs in an industrial environment. 

\begin{figure}[!t]
    \centering
    \includegraphics[width=0.6\textwidth]{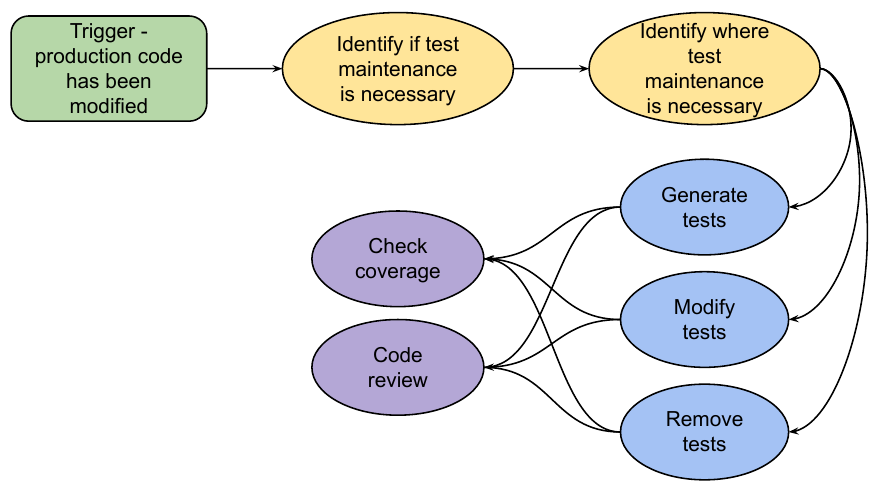}
    \caption{LLM agents could generate tests, or modify or remove existing tests, based on a combination of triggers, code coverage, and code quality review. Green = trigger, yellow = explored aspects, blue = new test maintenance actions, purple = new code analyses.}
    \label{fig:rq2_future_low}
\end{figure}

\smallskip\noindent\textbf{Use of Low-Level Triggers by LLMs (RQ2.1):} Because low-level triggers can be detected in the code or documentation, they could be directly used by LLM agents---potentially with other data signals---as the impetus to perform actions on the test suite. 

Figure~\ref{fig:rq2_future_low} illustrates a potential extension of our prototypes. In our prototypes, an LLM agent performed its own assessment of the code change. The triggers could be be blended with an LLM's analyses to increase the confidence in the action determined to be appropriate.  Once the need for test maintenance is predicted, additional agents could automatically modify the test suite. The low-level triggers  may suggest certain corresponding actions. Data extracted from other tools---e.g., code coverage information---or other analyses performed by LLMs---e.g., inspecting code quality---could be used along with the triggers to suggest  these modifications to the test suite. 

\begin{figure}[!t]
    \centering
    \includegraphics[width=0.8\textwidth]{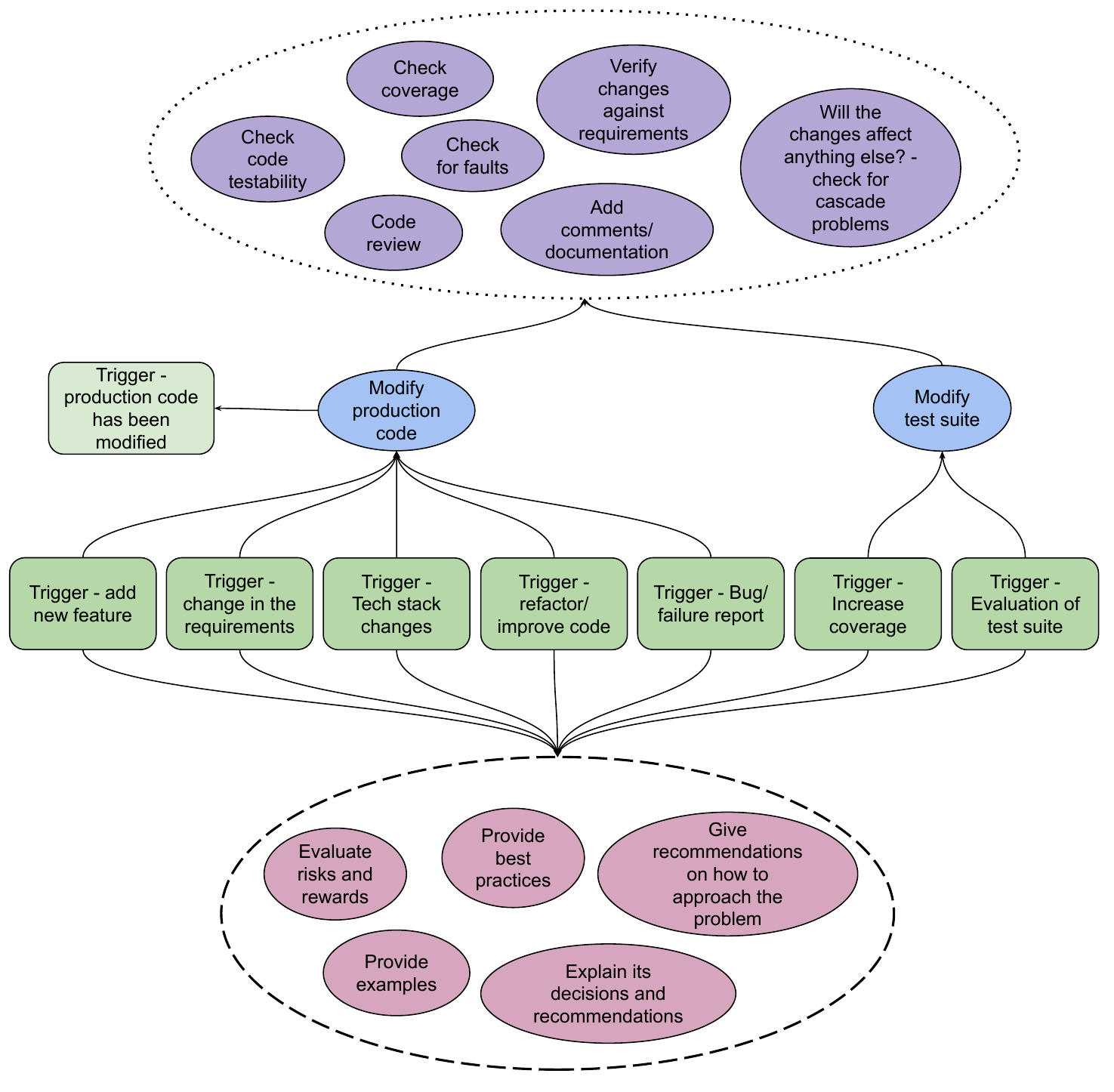}
    \caption{General actions an LLM agent could take based on a high-level trigger. Green = high-level trigger, light green = low-level trigger, purple = potential code or test quality actions, pink = general LLM actions, blue = code modification actions.}
    \label{fig:rq2_future_high_overview}
\end{figure}

\begin{coloredframe}{blue}
\textbf{Use of Triggers by LLMs (RQ2.1):} Low-level triggers, as well as other contextual information (e.g., coverage or quality analyses), could---among other tasks---identify and modify tests or suggest new tests. 
\end{coloredframe}

\smallskip\noindent\textbf{Use of High-Level Triggers by LLMs (RQ2.1):}  As high-level triggers reflect  development decisions, these triggers may not be detectable automatically. However, an alert that such a decision has been made could be used by LLM agents to perform or assist in planning and testing activities. Figure~\ref{fig:rq2_future_high_overview} suggests the types of actions that LLM agents could take based on high-level triggers, including general actions, code modifications, and other actions intended to improve test or source code quality. 

\begin{figure}[!t]
    \centering
    \includegraphics[width=0.8\textwidth]{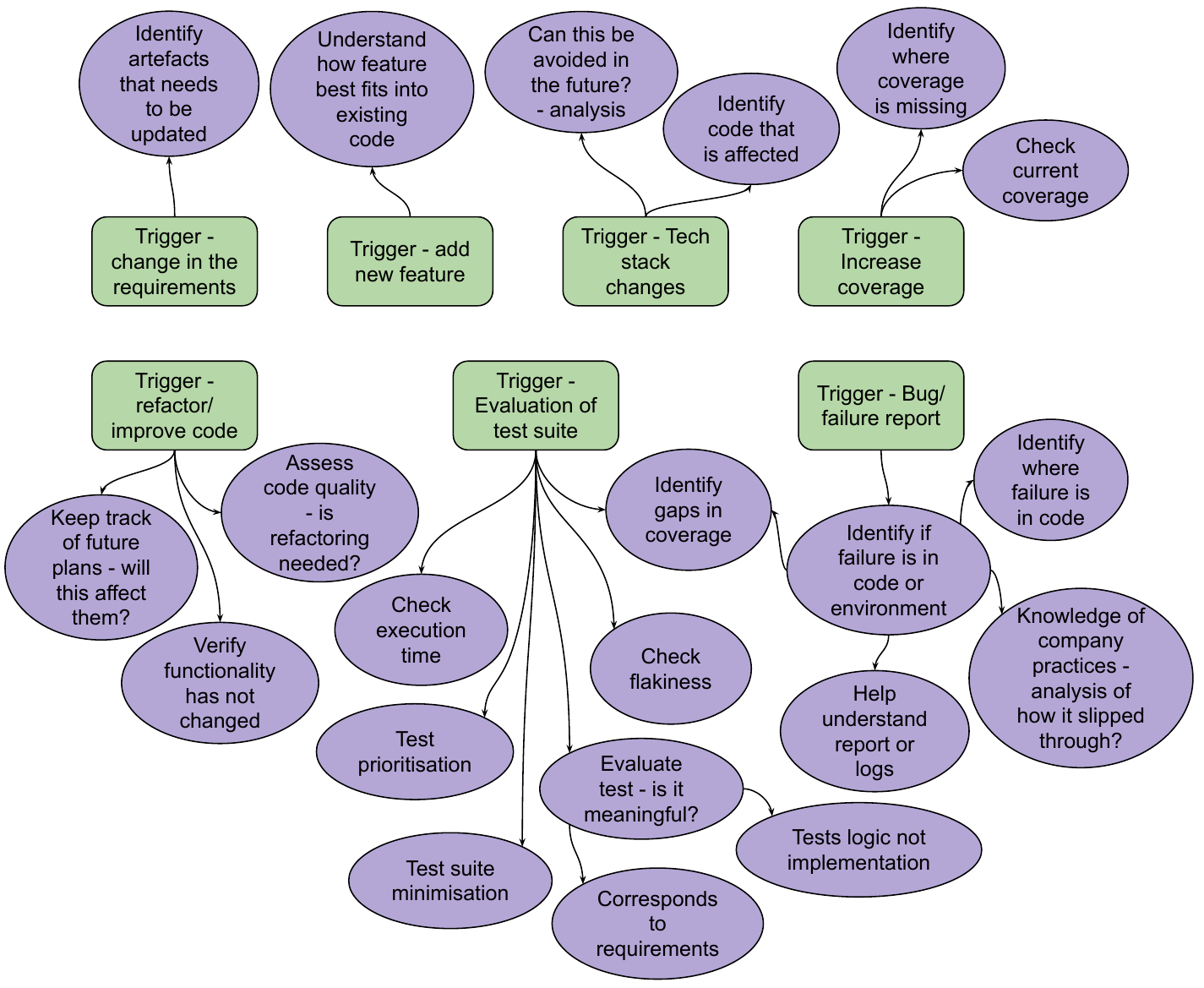}
    \caption{Suggestions on specific actions an LLM agent could take based on high-level triggers to improve the test suite quality. Green = high-level triggers, purple = potential LLM actions.}
    \label{fig:rq2_future_high_detail}
\end{figure}

Figure~\ref{fig:rq2_future_high_detail} offers examples of specific actions for each trigger, extracted from interview and survey responses and past literature. For example, given changed requirements, agents could identify the artifacts that need to be adjusted, including code and documentation (e.g., user stories). They could then recommend best practices for updating the code and test suite, assess risks and rewards of different solutions, modify the artifacts, generate tests that verify that the new requirements are met,  evaluate code quality, and suggest performance-improving refactoring to the code and test suite. 

\begin{coloredframe}{blue}
\textbf{Use of Triggers by LLMs (RQ2.1):} LLM agents could act on high-level triggers by performing or assisting in activities such as providing guidance, evaluating and offering recommendations on potential solutions, or modifying the source code or the test suite. 
\end{coloredframe}

\begin{comment}
        \item[RQ2.2] What potentially viable test maintenance actions could an LLM or LLM agent perform, based on these triggers?
        \item[RQ2.3] What considerations must be taken when implementing an LLM agent or other LLM-based system within an industrial environment?
    \end{description}
\end{comment}

\smallskip\noindent\textbf{Test Maintenance Actions (RQ2.2):} Interview and survey responses---as well as the literature review---suggest that LLMs have grat potential to perform or assist with test maintenance tasks such as: 
\begin{itemize}
    \item Establishing traceability between artifacts. 
    \item Helping developers understand where and why maintenance is needed.
    \item Generating, adapting, or optimizing test cases and suites.
    \item Explaining or summarizing code and tests. 
    \item Acting as an interactive testing partner, providing examples and advice.
    \item Assisting with general development activities that might appear during test maintenance, such as code review, analyzing and improving source and test code, and generating documentation.
\end{itemize}

Our prototypes demonstrate that LLM agents can summarize source code changes and test code, establish traceability between code changes and test code, and use those analyses to predict the need for test maintenance. Figures~\ref{fig:rq2_future_low}--\ref{fig:rq2_future_high_detail} further illustrate how triggers and other data could be used by LLM agents to perform additional actions. 

Significant research effort has been dedicated to code and test generation. However, it is important to also consider how LLMs can be applied to a broad range of test activities, as well as how LLMs and humans can interact as part of the test maintenance process.
%we also would like to emphasize the value of LLM agents for tasks beyond code generation, such as code review or acting as a testing partner. Test maintenance is more than the simple process of adding new tests for new code and updating tests when the code changes---it 
Test maintenance is a long and complex process intended to improve the quality of the project, consisting of more than simple adaptation of tests or creation of tests for new features.
%This process includes many tasks, such as assessing the strengths and limitations of the existing test suite, improving the efficiency of the test suite, and trying to improve the diversity of the scenarios tested. In particular, LLMs show great potential for tasks that blend source code and natural language, and researchers should explore a variety of these use cases within the context of the test maintenance process. 

\begin{coloredframe}{blue}
\textbf{Test Maintenance Actions (RQ2.2):} LLM agents can perform or assist with test maintenance activities, including---broadly---code generation or modification, acting as a conversational partner, and code analysis or documentation.  
\end{coloredframe}

\smallskip\noindent\textbf{Industrial Considerations (RQ2.3):} Our literature review, as well as the interview and survey responses, suggest that the following considerations are the most important when deploying LLMs in an industrial environment:
\begin{itemize}
    \item Except for code generation, there are few established benchmarks. Performance estimates may also be affected by data leakage, insufficient test suites, or poorly-defined prompts. Benchmarks should be taken as a basic, but not absolute, performance indicator. 
    \item In general, the larger the number of parameters, the better an LLM will perform without tuning. However, larger models carry significant computation requirements.  
    \item Deploying in-house offers control over the model, opportunities for tuning on organization data, and mitigates security and privacy risks. 
    \item Fine-tuning a model on organizational data can be  expensive and time-consuming. LLM agents may be a more efficient option, as RAG can offer access to current organizational data. %Since chunking data is a much faster task than training or tuning a model, RAG tools can be easily updated. Over time, the model could also be tuned or re-trained to attain even better performance. 
    \item Copyright and ownership of AI-generated output is a concern. Using models where the training data is public is a possible solution.
    \item Skepticism should be taken seriously. Expectations on LLM performance should be set carefully and LLM deployment should be done in a controlled manner to ensure that the models are sufficiently effective before all developers are encouraged to regularly use them.  
    %\item One of the main wishes that Ericsson employees had for support from LLMs was, naturally, generating and improving existing code. However, their test maintenance challenges were much more varied, as are LLM capabilities. Organizationally, LLM usage could benefit many different development and testing tasks, and these use cases should be widely explored.  
    \item Many test maintenance challenges are rooted in a lack of knowledge, both project-specific and general (e.g., on testing practices). To perform test maintenance, one may---for example---need to familiarize themselves with the source code, investigate the cause of a fault, or evaluate how code coverage is affected by code and test suite changes. Some of the most valuable capabilities LLMs can provide are summarization, insight, and recommendations regarding project artifacts. 
    \item LLMs should be deployed with some human oversight. Performance limitations and skepticism suggest that LLM output should be treated as ``recommendations'' reviewed by a human. LLMs also have value as conversational partners. LLMs should not fully replace human effort, but can augment the effectiveness and efficiency of human effort. 
\end{itemize}

\begin{coloredframe}{blue}
\textbf{Industrial Considerations (RQ2.3):} When deploying LLMs, organizations should consider model performance, whether to deploy in-house or externally, copyright and ownership, how to incorporate organizational data (e.g., fine tuning or RAG), how to establish and manage  expectations, which tasks LLMs should support, and how humans and LLMs should interact when performing these tasks.
\end{coloredframe}

\subsection{Proof-of-Concept Evaluation (RQ3)}\label{sec:discussion_rq3}

\noindent\textbf{Prototype Performance:} The prototype based on test summaries attains the best performance when test cases were updated in the ground truth, and worse performance when maintenance was not performed. The reason, in both cases, is that this prototype makes more positive predictions than the other prototype. Given the large improvement in performance in the former case, we recommend summarizing tests rather than working with raw code.

The initial performance of this prototype is promising, with relatively high recall. A comparable approach is DRIFT, proposed by Liu et al.~\cite{liu2023drift}. DRIFT achieved an average accuracy of 0.689, 0.712 for precision, 0.679 for recall, and 0.695 for F1 score. We cannot directly compare performance, as they used a different dataset and we were not able to re-implement DRIFT for the industrial codebase. However, we can make a rough comparison---our approach has a higher accuracy and equivalent recall, but lower precision. 

The primary factor negatively affecting performance of our prototype is low precision---the prototype yields is more likely to falsely predict that maintenance is required than to miss a test that actually requires maintenance. However, recall is arguably more important than precision for this task, as long as the user is not inundated with too many FP results.

Future work should consider how to improve recall even further, as well as how to improve precision without a loss in recall. Potential paths to better performance include newer and larger models, consideration of models tuned for code tasks, and fine-tuning the model on the Ericsson codebase. We also used a single embedding model, rather than selecting different embedding models for different tasks. The similarity measurement used by the embedding model can impact  performance. We can also consider prompt engineering techniques such as query expansion~\cite{jagerman2023queryexpansionpromptinglarge}.  %Query expansion could also be used to reformulate prompts and tool input---e.g., adding synonyms---to attain more relevant results~\cite{jagerman2023queryexpansionpromptinglarge}. 

\begin{coloredframe}{blue}
\textbf{Prototype Performance (RQ3):} Our prototype based on test summaries has a high recall, showing its potential for this task. However, its precision is relatively low, and could be improved in future work through larger or specialized models, fine-tuning, prompt engineering, and alternative embedding models. 
\end{coloredframe}

\smallskip\noindent\textbf{Developer Assessment:} In both developer validations, the participants showed interest in the prototype as a ``second opinion'' on their own judgment, even if the performance was not perfect. The prototype could be executed more quickly than the test suite, and offered an additional data signal that could be incorporated into the existing maintenance workflow.

In the second validation, the developer of the industrial codebase provided additional insight into the quantitative performance results. The developer found that several of the ``false positives'' were actually correct---either tests that the developer missed initially when performing maintenance, or cases where the ground truth was incorrectly captured. Further, in several cases, even an inaccurate suggestion can offer some benefit. The prototype does not just state the names of test cases, but explains its rationale, indicates its confidence, and may make additional suggestions. This output can offer insight or suggest additional tests to create. 

%Currently, to increase explainability, the prototypes issue output on the result of each step performed during the process of arriving at the final recommendation. The participants expressed that the output was too verbose, and that they would prefer a shorter trace coupled with a better explanation during the final output.
%They also requested suggestions on how to change the test cases. In addition, both participants stressed that the prototypes must be integrated with existing development tools, e.g., as an IDE plug-in, or at least easy to trigger from these tools. 

\begin{coloredframe}{blue}
\textbf{Developer Assessment (RQ3):} The prototype was considered useful, as it is faster to execute than the test suite and offers additional input for test maintenance decisions. Some of the suggested tests were missed by the developer during the original maintenance process, and the suggestions made by the prototype may be useful, even if not fully correct.
\end{coloredframe}

\subsection{Threats to Validity}
\label{sec:discussion_threats}

\noindent\textbf{External Validity:} We conducted a case study at one company, and the prototype was evaluated on a single project. However, we believe that our exploratory findings will generalize, as well as our qualitative findings for RQ3. Our findings in RQ1--2 were primarily derived from literature, then validated in the company context. The prototype architecture is not closely coupled to Ericsson. However, future work should extend beyond Ericsson. %Therefore, the core ideas guiding the design of the prototypes and suggestions regarding improved performance should hold outside of our evaluation. 

\smallskip\noindent\textbf{Internal Validity:} We draw conclusions based on a limited sample that may not be fully representative of the domain. When conducting surveys, we lack control over the response rate and there can be subjectivity in interpreting quantitative answers~\cite{ghazi2018survey}. Interpretation of qualitative data  may also be biased. To mitigate these threats, we performed data triangulation, validating findings across multiple data sources~\cite{runeson2009guidelines}.

When creating the evaluation dataset, we assumed that tests were updated because of the source changes in the same commit. There may also be cases where tests were changed in an earlier or later commit, leading to false positives that were actually valid suggestions. As a result, prototype performance may be lower than the potential real-world performance. %Further, ground truth is known for each commit, not for each code change. To assess accuracy, we merge results for all code changes for a commit. This could lead to some noise in the assessment of results. However, we speculate that the dataset is sufficiently accurate to offer an indication of the capabilities of the prototypes. 

\smallskip\noindent\textbf{Construct Validity:} There is a risk of participants misunderstanding or misinterpreting questions. This was mitigated by running pilot trials. The wording of some survey and interview questions was clarified after the pilot.

%The prototypes have not undergone rigorous verification, and may contain faults. This threat was partially mitigated by using open-source functionality for LLM agents and RAG tools from the LangChain framework~\cite{langchain}.

Some source and test code was excluded from the dataset. These exclusions could introduce noise, e.g., tests may be included in the ground truth that were connected to excluded source code, or vice-versa. However, we manually reviewed all exclusions. 

\section{Conclusions}\label{sec:conclusions}
In this study, we explore the capabilities of LLMs and LLM agents to support test maintenance. We have identified 44 low and high-level triggers that can indicate a need for test maintenance. Based on these triggers and other data signals, we speculate that LLMs and LLM agents could perform or assist with test maintenance activities, including code generation or modification, acting as a conversational partner, and code analysis or documentation.    

% We identified 37 low-level changes to code and documentation that can trigger the need for test maintenance. However, such a need only occurs in situations where test cases exist that invoke the changed code, then cause and detect changes to the behavior of that code. Future work should augment triggers with additional contextual information (e.g., coverage or code quality analyses) to increase the accuracy and complexity of the actions taken by LLMs. 

% We also identified seven high-level development decisions that frequently trigger a need for test maintenance, including adding new features, requirements changes, fault discovery, increasing the coverage of a test suite, refactoring, tech stack changes, and test suite evaluation. LLM agents could act on such triggers by, e.g., providing examples or best practices, evaluating potential solutions, offering recommendations on how to improve potential solutions, modifying source code, or modifying the test suite. 

When deploying LLMs, organizations should consider model performance, whether to deploy in-house or externally, copyright and ownership, how to incorporate organizational data (e.g., fine tuning or RAG), how to establish and manage  expectations, which tasks LLMs should support, and how humans and LLMs should interact when performing these tasks.

Our prototype based on test summaries has a high recall, demonstrating that LLM agents could play a practical role in test maintenance. When inspecting a subset of false positives from the prototype, we discovered that some of these tests were missed by the developer during the original maintenance process, and that the suggestions made by the prototype may be useful, even if not fully correct. Still, future research should consider how the precision of the prototype could be improved. Other test maintenance activities should also be explored. Collectively, these contributions advance our theoretical and practical understanding of how LLMs can be deployed to benefit industrial test maintenance processes.

\section{Acknowledgements}

Support for this research was provided by the Wallenberg AI, Autonomous Systems and Software Program (WASP) funded by the Knut and Alice Wallenberg Foundation, as well as Software Center project 63 ``AI-Enabled Test Automation, Generation, and Optimization''.

\bibliographystyle{elsarticle-num-names}
\bibliography{main}

\clearpage 

%\pagenumbering{gobble}

%\appendix 
%\input{prompts}

\end{document}